\documentclass[conf]{new-aiaa}
\usepackage[titles]{tocloft}
\usepackage[utf8]{inputenc}
\usepackage[table,xcdraw]{xcolor}
\usepackage{graphicx}
\usepackage{amsmath}
\usepackage{listings}
\usepackage{color}
\usepackage{float}
\usepackage[version=4]{mhchem}
\usepackage{siunitx}
\usepackage{longtable,tabularx}
\usepackage{cleveref}
\usepackage[ruled,vlined]{algorithm2e}
\usepackage{multirow}
\usepackage{subcaption}

\definecolor{dkgreen}{rgb}{0,0.6,0}
\definecolor{gray}{rgb}{0.5,0.5,0.5}
\definecolor{mauve}{rgb}{0.58,0,0.82}
\definecolor{htmlfe0000}{RGB}{254,0,0}
\definecolor{html34ff34}{RGB}{52,255,52}
\definecolor{htmlfd6864}{RGB}{253,104,100}
\definecolor{html32cb00}{RGB}{50,203,0}

\lstdefinelanguage{Scheme}{
  morekeywords=[1]{define, define-syntax, define-macro, lambda,
    define-stream, stream-lambda},
  morekeywords=[2]{begin, call-with-current-continuation, call/cc,
    call-with-input-file, call-with-output-file, case, cond,
    do, else, for-each, if,
    let*, let, let-syntax, letrec, letrec-syntax,
    let-values, let*-values,
    and, or, not, delay, force,
    quasiquote, quote, unquote, unquote-splicing,
    map, fold, syntax, syntax-rules, eval, environment, query,
    display, newline,list, apply, null?, car, cdr, for-each, 
make-vector, vector-length, vector-ref, vector-set!, eqv?, eq?, equal?, set!, 
define-record-type, fields, mutable, immutable, assert, parent, with-exception-handler,
  },
  morekeywords=[3]{import, export},
  alsoletter=?!-,
  alsodigit=\$\%&*+./:<=>@^_~,
  sensitive=false,
  morecomment=[l]{;},
  morecomment=[s]{/*}{*/},
  morestring=[b]",
  basicstyle=\small\ttfamily,
  keywordstyle=\bf\ttfamily\color[rgb]{0,.3,.7},
  commentstyle=\color[rgb]{0.133,0.545,0.133},
  stringstyle={\color[rgb]{0.75,0.49,0.07}},
  upquote=true,
  breaklines=true,
  breakatwhitespace=true,
  literate=*{`}{{`}}{1}
}

\title{Peaking into the Black-box: Prediction Intervals Give Insight into Data-driven Quadrotor Model Reliability}

\author{J.J. van Beers \footnote{j.j.vanbeers@tudelft.nl, PhD candidate, Faculty of Aerospace Engineering: Control and Simulation} and C.C. de Visser \footnote{Associate Professor, Faculty of Aerospace Engineering: Control and Simulation}}
\affil{Delft University of Technology, 2629 HS Delft, The Netherlands}

\begin{document}

\renewcommand*\cftsecnumwidth{2.5em}
\renewcommand*\cftsubsecnumwidth{2.5em}

\maketitle

\begin{abstract}
    Ensuring the reliability and validity of data-driven quadrotor model predictions is essential for their accepted and practical use. This is especially true for grey- and black-box models wherein the mapping of inputs to predictions is not transparent and subsequent reliability notoriously difficult to ascertain. Nonetheless, such techniques are frequently and successfully used to identify quadrotor models. Prediction intervals (PIs) may be employed to provide insight into the consistency and accuracy of model predictions. This paper estimates such PIs for polynomial and Artificial Neural Network (ANN) quadrotor aerodynamic models. Two existing ANN PI estimation techniques - the bootstrap method and the quality driven method – are validated numerically for quadrotor aerodynamic models using an existing high-fidelity quadrotor simulation. Quadrotor aerodynamic models are then identified on real quadrotor flight data to demonstrate their utility and explore their sensitivity to model interpolation and extrapolation. It is found that the ANN-based PIs widen considerably when extrapolating and remain constant, or shrink, when interpolating. While this behaviour also occurs for the polynomial PIs, it is of lower magnitude. The estimated PIs establish probabilistic bounds within which the quadrotor model outputs will likely lie, subject to modelling and measurement uncertainties that are reflected through the PI widths.
\end{abstract}

\section{Introduction}

\lettrine{D}{ata-driven} system identification techniques are often employed to develop models of highly non-linear systems, such as the quadrotor. These data-driven approaches to quadrotor aerodynamic model identification are convenient and attractive due to limited knowledge on analytical descriptions of the quadrotor \cite{Sun2018_GrayBox,powers2013_AeroInfluence}. Subsequently, many quadrotor models have been successfully identified in literature using a plethora of grey and black-box system identification techniques \cite{Sun2018_GrayBox,Bansal2016,bauersfeld2021neurobem}.

Of particular note are the use of polynomial step-wise regression and Artificial Neural Networks (ANNs). Sun et al. \cite{Sun2018_GrayBox} construct polynomial models of the quadrotor for the high-speed ($14$ $ms^{-1}$) flight regime through step-wise regression. By exploiting knowledge on influential states of the system, informed choices on the candidate regressors are made. Subsequently, the stepwise algorithm sequentially selects the best model structure, hence the term 'grey'-box. Even though the selected regressors are visible in the final model, their selection is inherently phenomenological. Hence, these models occasionally produce unexpected predictions, particularly when noise is present in the system. ANNs are an increasingly popular modelling choice by virtue of their generalizability and aptitude for capturing unknown non-linearities \cite{Bansal2016}. Consequently, they are employed to approximate dynamics in unexplored and extreme regions of the flight envelope. Take, for instance, the use of ANNs to facilitate quadrotor model identification on high-speed flight data (up to $18$ $ms^{-1}$) by Bauersfeld et al. \cite{bauersfeld2021neurobem}. However, as a black-box modelling technique, it is ambiguous as to how exactly the ANN inputs (e.g. system states) interact to produce the model outputs. Evaluating the reliability and validity of the resultant ANN models is challenging and proves to be a potential barrier to their widespread use in the aerospace industry \cite{Schumann03onverification}. Indeed, uncertainty in the inputs may culminate in some unexpected model outputs for similar (e.g. noisy) inputs \cite{Khosravi2011_NN_Prediction_intervals}. Ensuring safety and reliability are paramount to the accepted use of autonomous quadrotors and, thus, this lack of model reliability measures for such grey and black-box models needs to be addressed. 

One way to describe the reliability of a model's prediction is through its confidence in that prediction. To this end, the associated prediction intervals (PIs) - the interval wherein a future observation will lie with a given probability (i.e. confidence) - may be used as a proxy for reliability. PIs account for both variations due to model uncertainty and variations due to measurement uncertainty (e.g. noise), up to a specifiable confidence level (typically 95\%). Therefore, the often-used confidence interval (CI), which describes only model uncertainty, is a component of the PI.

Fortunately, for polynomial models, there are valid analytical formulations for obtaining these PIs. Although, these have yet to be applied to the aforementioned quadrotor grey-box models. In contrast, due to the inherent obscurity behind black-box models, analytical formulations that maintain the validity of any estimated PIs are challenging, if not impossible, to derive. Consequently, some ANN literature \cite{Khosravi2011_NN_Prediction_intervals,KhosraviLUBE,pearce2018highquality,pearce2020uncertainty} propose various methods to estimate these PIs. One of the more straightforward and convenient approaches is known as the bootstrap method wherein an additional ANN is employed to estimate the PIs \cite{Khosravi2011_NN_Prediction_intervals}. This approach is particularly attractive as it may be applied to an already trained ensemble of ANNs without needing to retrain the underlying ensemble. In their review of popular ANN PI estimation techniques, Khosravi et al. \cite{Khosravi2011_NN_Prediction_intervals} found that the bootstrap approach did well to reflect uncertainty in the inputs. However, the bootstrap method often produces excessively wide PIs, which is undesirable. Instead, it is perhaps more lucrative to augment the training procedure in some way to ensure high-quality (i.e. valid, yet narrow) PIs. Accordingly, Pearce et al. \cite{pearce2018highquality} propose a quality-driven direct estimation of the PIs which incorporates common PI quality metrics in the cost function. This approach builds upon the lower upper bound estimation (LUBE) \cite{KhosraviLUBE} method, but is argued to produce higher quality PIs while maintaining compatibility with common ANN libraries \cite{pearce2018highquality}. These ANN PI estimation methods, or indeed any others, provide a clear utility in establishing model reliability but have yet to be applied to ANN quadrotor aerodynamic models.

The contributions of this paper are two-fold; first, the aforementioned PI estimation methods are numerically validated for quadrotor aerodynamic models for the first time through an existing high-fidelity quadrotor simulation developed by Sun et al. \cite{Sun_ControlDoubleFailure,Sun2018_GrayBox}. Subsequently, to further demonstrate their utility, models and validated PIs are applied to real high-speed (up to $14$ $ms^{-1}$) quadrotor flight data. To the best of our knowledge, this is the first time such PIs are identified for quadrotor models. How the PIs respond when tasked with model interpolation and extrapolation is also explored. Through these analyses, insights into the suitability and limitations of the all PI estimation methods are uncovered. 

The remainder of this paper is structured as follows: \cref{sec:quadModelInputs} provides some context on the quadrotor and defines important states for model identification. Further background on prediction intervals and the considered estimation techniques are presented in \cref{sec:prediction_intervals}. Thereafter, the numerical validation procedure and results are given in \cref{sec:numerical_validation}. Finally, \cref{sec:application_real_quad_data} applies the PI estimation techniques to real quadrotor flight data and evaluates the sensitivity of the estimated PIs to model interpolation and extrapolation.

\section{Derivation of Quadrotor Model Inputs}\label{sec:quadModelInputs}

As this paper is concerned with the validity of PIs for quadrotor models, it is important to first provide context on the system itself. This then motivates the choices for the necessary inputs variables of the aerodynamic models.

\begin{figure}[!b]
    \begin{minipage}{.48\textwidth}
        \centering
        \includegraphics[width = \textwidth]{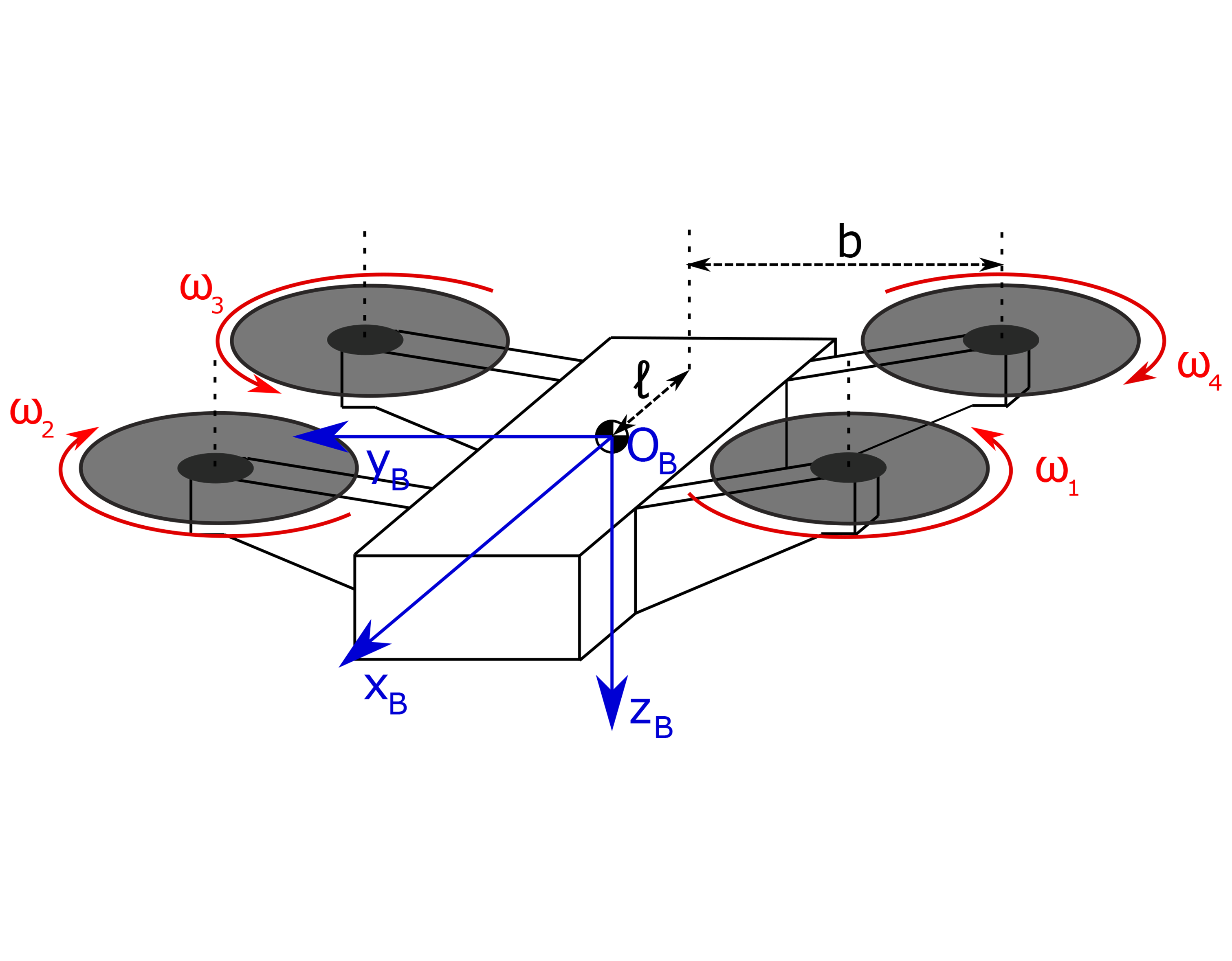}
        \caption{Quadrotor body reference frame.}
        \label{fig:QuadReferenceFrame}
    \end{minipage}%
    \hfill
    \begin{minipage}{.48\textwidth}
        \centering
        \includegraphics[width=\textwidth]{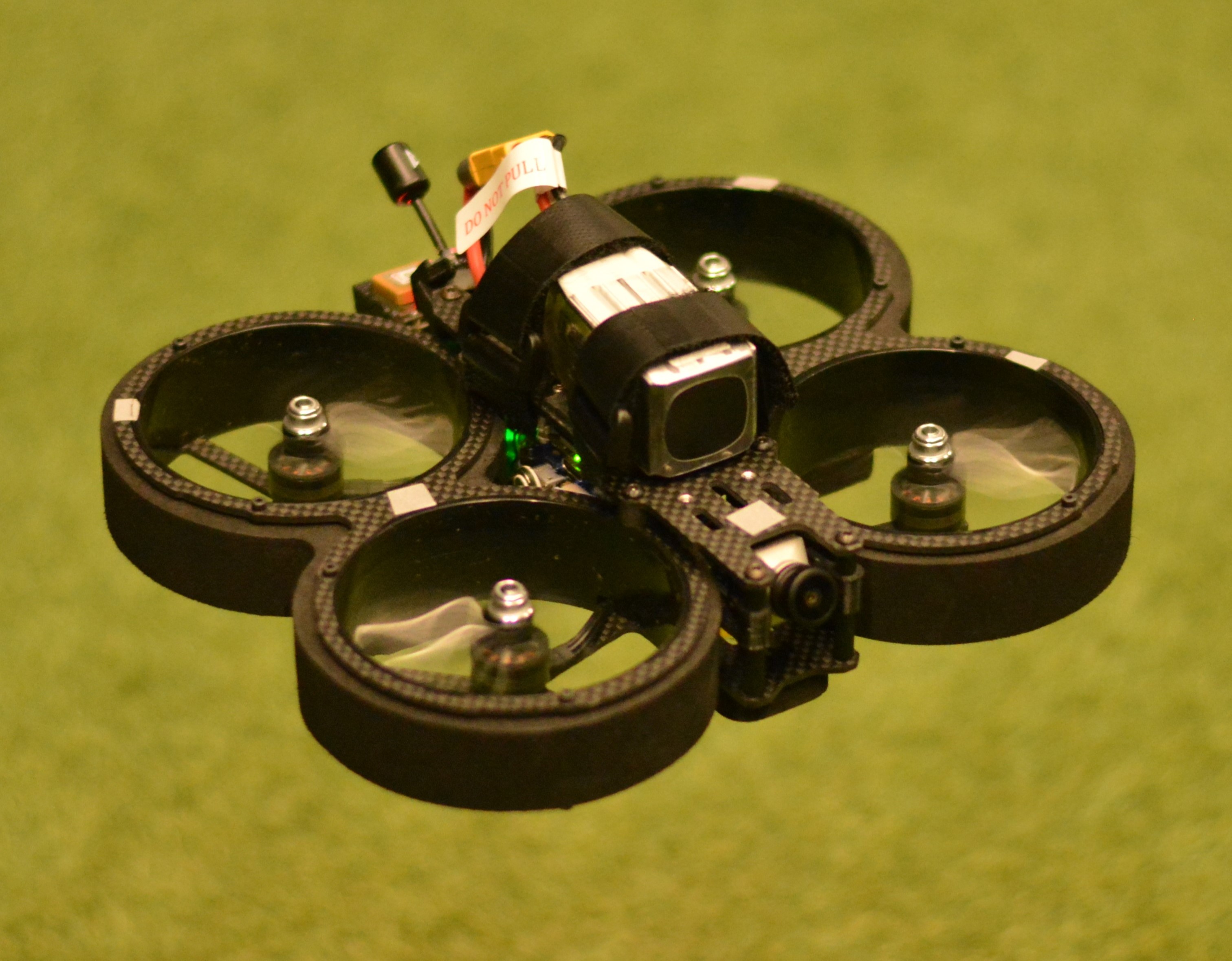}
        \caption{Quadrotor test platform, the \textit{HDBeetle}.}
        \label{fig:HDBeetle}
    \end{minipage}%
\end{figure}

Following Newton's laws of motion, \cref{eq:QuadSimple_Force_B} and \cref{eq:QuadSimple_Moment_B} can be derived to describe the kinematics of the quadrotor in the body reference frame, $\{B\}$ (see \cref{fig:QuadReferenceFrame}). Here, $m$ denotes the mass of the quadrotor and $I_{v}$ its moment of inertia. $\mathbf{V}_{B} = [u, v, w]^{T}$ represents the body velocity and $\mathbf{\Omega}_{B} = [p, q, r]^{T}$ the body rotational rates. $\mathbf{F}_{B}$ and $\mathbf{M}_{B}$ describe the (aerodynamic) forces and moments acting on the quadrotor body. The rotational matrix from the North-East-Down (NED) earth reference frame, $\{E\}$, to $\{B\}$ is denoted by $R_{BE}$. $\mathbf{g}$ is the gravitational vector in $\{E\}$.

\begin{equation}\label{eq:QuadSimple_Force_B}
    m \left ( \dot{\mathbf{V}}_{B} + \boldsymbol{\Omega}_{B} \times \mathbf{V}_{B} \right ) = R_{BE}m\mathbf{g} + \mathbf{F}_{B}
\end{equation}

\begin{equation}\label{eq:QuadSimple_Moment_B}
    \mathbf{I_{v}}\dot{\boldsymbol{\Omega}}_{B} + \boldsymbol{\Omega}_{B} \times \mathbf{I_{v}}\boldsymbol{\Omega}_{B} = \mathbf{M}_{B}
\end{equation}

Subsequently, aerodynamic models of the quadrotor should incorporate the body velocity, $\mathbf{V}_{B}$, and the body rotational rates, $\mathbf{\Omega}_{B}$. 

\begin{equation}\label{eq:rotor_thrust}
    T = \kappa _{0} \sum \omega _{i} ^{2}
\end{equation}

Analytical models of the quadrotor provide further insight into useful input variables. For example, \cref{eq:rotor_thrust} describes a simple model of the total thrust generated by the rotors \cite{Svacha2017_QuadAerodynamics,powers2013_AeroInfluence}. $\omega _{i}$ denotes the angular velocity of the $i^{\text{th}}$ rotor\footnote[2]{It is assumed that the rotors are aligned with the $x_{B}$-$y_{B}$ plane.} and $\kappa _{0}$ is a constant for a given quadrotor which may be identified from hovering data. Note that, per the definition of $\{B\}$ in \cref{fig:QuadReferenceFrame}, the thrust vector points in the negative $z_{B}$ direction. 

\begin{equation}\label{eq:avg_omega}
    \omega_{avg} = \sqrt{\frac{\sum_{i=1}^{4}{\omega^{2}_{i}}}{4}}
\end{equation}

Consequently, the rotor speeds, $\mathbf{\omega}$, are informative for aerodynamic quadrotor models. However, to avoid over-fitting, their aggregated influence (through \cref{eq:avg_omega} \cite{Sun2018_GrayBox}) is used instead. Likewise, the differential thrust between the rotors can account for differences between the rotors. Following the quadrotor layout in \cref{fig:QuadReferenceFrame}, the so-called control rolling, pitching, and yawing moments are defined in \cref{eq:control_moment_Up}, \cref{eq:control_moment_Uq}, and \cref{eq:control_moment_Ur} respectively. Here, $S_{R} \in \{-1, 1\}$ describes the rotation direction of the first rotor (such that clockwise = 1). 

\begin{equation}\label{eq:control_moment_Up}
    U_{p} = \left ( {\omega_{1}} + {\omega_{4}} \right ) - \left ( {\omega_{2}} + {\omega_{3}} \right )
\end{equation}

\begin{equation}\label{eq:control_moment_Uq}
    U_{q} = \left ( {\omega_{1}} + {\omega_{2}} \right ) - \left ( {\omega_{3}} + {\omega_{4}} \right )
\end{equation}

\begin{equation}\label{eq:control_moment_Ur}
    U_{r} = S_{R} \left [ \left ( {\omega_{1}} + {\omega_{3}} \right ) - \left ( {\omega_{2}} + {\omega_{4}} \right ) \right ]
\end{equation}

Furthermore, literature on analytical quadrotor models \cite{Svacha2017_QuadAerodynamics,powers2013_AeroInfluence} also emphasize the importance of the angle of attack of the rotor plane, the influence of the induced velocity, and the effects of blade flapping on the aerodynamics of the quadrotor. While the angle of attack itself is difficult to obtain accurately for the quadrotor, the attitude angles (i.e. $[\phi, \theta, \psi]^{T}$) may instead be used as a proxy of this, assuming that $|\theta| < \frac{\pi}{2}$ and $|\phi| < \frac{\pi}{2}$. The quadrotor attitudes are represented by their trigonometric identities (i.e. $\sin{(\cdot)}$ and $\cos{(\cdot)}$) to accommodate their periodic nature. This also bounds, and effectively normalizes, the attitude. The induced velocity describes the added velocity of the airflow as it passes through the rotors and may be computed through \cref{eq:induced_velocity} where $v_{h}$ denotes the induced velocity during hover \cite{powers2013_AeroInfluence}. 

\begin{equation}\label{eq:induced_velocity}
    v_{in} = \frac{v_{h}^{2}}{\sqrt{\left ( V\cos\alpha _{r} \right )^{2} + \left ( v_{in} - V\sin\alpha _{r} \right )^{2}}}
\end{equation}

The effects of blade flapping are related to the advance ratios \cite{powers2013_AeroInfluence} - a variable which contextualizes the body velocities with respect to the rotor tip velocities. \Cref{eq:advance_ratios} describes how these advance ratios may be computed. 

\begin{equation}\label{eq:advance_ratios}
    \begin{array}{cccc}
        \mu_{x} = \frac{u}{\omega_{avg} R} \quad, & \mu_{y} = \frac{v}{\omega_{avg} R} \quad, & \mu_{z} = \frac{w}{\omega_{avg} R} \quad, & \mu_{v_{in}} = \frac{w}{\omega_{avg} R}
    \end{array}
\end{equation}

Therefore, the basis of input states for the quadrotor aerodynamic models identified in this paper are summarized in \cref{eq:state_vector}.

\begin{equation}\label{eq:state_vector}
    X = \left [  {u}, {v}, {w}, {v_{in}}, {p}, {q}, {r}, \omega_{avg}, \sin{\phi}, \cos{\phi}, \sin{\theta}, \cos{\theta}, {U}_{p}, {U}_{q}, {U}_{r}, \mu_{x}, \mu_{y}, \mu_{z}, \mu_{v_{in}} \right ]
\end{equation}
\section{Prediction Interval Estimation Methods}\label{sec:prediction_intervals}

Although regression aims to approximate a dependent variable given a set of predictors, measurements of these are often contaminated with bias and noise. Further uncertainties and errors arise from modelling misspecification. Both sources of uncertainty induce a variation in model predictions, which may be modelled through \cref{eq:total_variance} \cite{Khosravi2011_NN_Prediction_intervals}. Here, $\boldsymbol{\sigma}_{\hat{\boldsymbol{\epsilon}}}^{2}$ encompasses the uncertainty associated with the measurements (i.e. dependent variable) and $\boldsymbol{\sigma}_{\hat{\mathbf{y}}}^{2}$ denotes the variance due to modelling misspecification and uncertainty. The total variance, $\boldsymbol{\sigma}^{2}$, may then be used to construct the prediction intervals (PIs) for a specifiable confidence level, $1-\alpha$. Typically, $\alpha = 0.05$ is used, leading to $95\%$ intervals. Note that, since the PIs consider both $\boldsymbol{\sigma}_{\hat{\boldsymbol{\epsilon}}}^{2}$ and $\boldsymbol{\sigma}_{\hat{\mathbf{y}}}^{2}$, they encompass the associated confidence intervals (CIs). 

\begin{equation}\label{eq:total_variance}
    \boldsymbol{\sigma}^{2} = \boldsymbol{\sigma}_{\hat{\mathbf{y}}}^{2} + \boldsymbol{\sigma}_{\hat{\boldsymbol{\epsilon}}}^{2}
\end{equation}

The corresponding PI bounds, for a given measurement value, $y_{i}$, and accompanying total variance, $\sigma _{i}^{2}$, can be obtained through \cref{eq:PI_bounds}. In \cref{eq:PI_bounds}, $t^{1-\alpha/2}_{N-2}$ gives two sided test statistic for the chosen confidence level, $1 - \alpha$, and $n$ gives the degrees of freedom (i.e. number of observations of the sample, $i$).

\begin{equation}\label{eq:PI_bounds}
\begin{array}{ccc}
    L_{i} = \hat{y}_{i} - t^{1-\alpha/2}_{N-2}\frac{\sigma _{i}}{\sqrt{n}}, & & U_{i} = \hat{y}_{i} +  t^{1-\alpha/2}_{N-2}\frac{\sigma _{i}}{\sqrt{n}}
\end{array}
\end{equation}


The PI coverage probability (PICP), defined in \cref{eq:PICP}, accesses the validity of a PI by verifying if the desired confidence level is reached \cite{Khosravi2011_NN_Prediction_intervals,pearce2018highquality}. $N$ denotes the number of test data points and $c_{i}$ describes if the measurement is contained within the bounds of the PI (where $L_{i}$ = lower bound, $U_{i}$ = upper bound). Thus, PIs are considered valid if $PICP \ge 1 - \alpha$. Note that the PICP metric alone imposes no requirement on the width of the PIs, which may be impractically wide. 



\begin{equation}\label{eq:PICP}
\begin{array}{cccccc}
    PICP = \frac{1}{N}\sum_{i=1}^{N}c_{i} &, & \text{where:} & & 
    c_{i} = \left \{ \begin{array}{ll} 1, & y_{i} \in [L_{i}, U_{i}] \\ 0, & \text{otherwise}\end{array} \right.
\end{array}
\end{equation}

Therefore, complimentary to the PICP, the mean PI width (MPIW) is often used to parameterize the width of the PIs and is typically normalized with respect to the range of the measurement data. The MPIW is given by \cref{eq:MIPW}. Both the PICP and the MPIW should be used in tandem to describe the overall PI quality, seeing as the deficiencies of one metric are perfectly accommodated by the other \cite{KhosraviLUBE}. A high quality PI is thus valid, yet narrow. 

\begin{equation}\label{eq:MIPW}
    MIPW = \frac{1}{N}\sum_{i=1}^{N}{(U_{i} - L_{i})}
\end{equation}

\subsection{Polynomial}
Conveniently, for polynomial models, the total variance associated with a prediction may calculated through \cref{eq:OLS_sigma} under the assumptions that the measurement error term, $\epsilon$, is zero-mean, independent, and identically distributed. In \cref{eq:OLS_sigma}, $\mathbf{x_{0}}$ denotes the input data regressor matrix for which predictions should be made (i.e. to predict $y_{0}$), $\mathbf{X}$ represents the regressor covariance matrix used for training the model, $\mathbf{I}$ gives the identity matrix and $\sigma_{e}^{2}$ may be approximated through \cref{eq:sigma_est}.


\begin{equation}\label{eq:OLS_sigma}
    \hat{\sigma}_{0}^{2} = \sigma_{e}^{2} \left ( \mathbf{I} + \mathbf{x}_{0} \left ( \mathbf{X}^{T} \mathbf{X} \right )^{-1} \mathbf{x}_{0}^{T} \right )
\end{equation}

\begin{equation}\label{eq:sigma_est}
    \hat{\sigma}_{e}^{2} = \frac{1}{N_{t}-2}\sum_{i=1}^{N_{t}}{\hat{e}_{i}^{2}}
\end{equation}

In \cref{eq:sigma_est}, $N_{t}$ represents the number of samples used for training and $\hat{e}_{i} = y_{i} - \hat{y}_{i}$ denotes the residual error for training sample, $i$. The computed $\hat{\sigma} _{0}$ may then be inputted into \cref{eq:PI_bounds} to construct the polynomial PI.

\subsection{ANN: Bootstrap}

The bootstrap method exploits the inherent stochasticity of the ANN training process and can be applied to an already trained ensemble of ANNs. The spread of predictions across an ensemble of ANN models, initialized with distinct weights and trained on different subsets of the training data set, gives an approximation of the model misspecification uncertainty \cite{Khosravi2011_NN_Prediction_intervals,pearce2018highquality}. The associated variance may be constructed through \cref{eq:bootstrap_sig_hat} \cite{Khosravi2011_NN_Prediction_intervals}, where $B$ gives the number of ANN models, $\hat{y}_{i}$ denotes the ensemble averaged prediction of the measurement. For this reason, the bootstrap method can only be applied to an ensemble of ANNs. Although the appropriate number of ensembles depends on the system, increasing the number of ensembles does not always lead to improved performance \cite{Khosravi2011_NN_Prediction_intervals}. 

\begin{equation}\label{eq:bootstrap_sig_hat}
    \hat{\sigma}_{y_{i}}^{2} = \frac{1}{B-1}\sum_{j=1}^{B}{\left ( \hat{y}_{i}^{j} - \hat{y}_{i} \right )^{2}}
\end{equation}

The variance due to the measurement errors, $\sigma_{\epsilon_{i}}^{2}$, is typically unknown and cannot easily be obtained explicitly. Instead, an additional ANN may be used to (implicitly) estimate this variance. Through the residuals between the model predictions and the measurement data, a new data set, $D_{r} = \{ x_{i}, r_{i}^{2} \}_{i=1}^{m}$, may be constructed linking the inputs, $x_{i}$, to the variance residuals, $r_{i}^{2}$, defined in \cref{eq:bootstrap_variance_residuals} \cite{Khosravi2011_NN_Prediction_intervals}. Note, that the data used to derive $D_{r}$ should be distinct to that used for training the initial ensemble of ANNs. 

\begin{equation}\label{eq:bootstrap_variance_residuals}
    r_{i}^{2} = \max \left ( \left ( y_{m,i} - \hat{y}_{i})^{2} - \hat{\sigma}_{y_{i}}^{2} \right ), 0 \right )
\end{equation}

The $\sigma_{\epsilon_{i}}^{2}$-predictor ANN may then be used to approximate $\sigma_{\epsilon_{i}}^{2}$ by minimizing the cost function defined in \cref{eq:bootstrap_cost_function} \cite{Khosravi2011_NN_Prediction_intervals}. Note that this cost function is derived assuming that $\sigma_{\epsilon_{i}}^{2}$ is Gaussian. Moreover, due to the approximation of $\sigma_{\epsilon_{i}}^{2}$ through the residual, $r_{i}^{2}$, the Bootstrap method tends to estimate wider PIs \cite{Khosravi2011_NN_Prediction_intervals}. 


\begin{equation}\label{eq:bootstrap_cost_function}
    J = \frac{1}{2}\sum_{i=1}^{m}{\left [ \ln{(\sigma_{\epsilon_{i}}^{2})} + \frac{r_{i}^{2}}{\sigma_{\epsilon_{i}}^{2}} \right ]}
\end{equation}

\subsection{ANN: Quality Driven}
Instead of estimating the PIs after-the-fact, another approach is to estimate them directly \cite{KhosraviLUBE}. This is advantageous in that the PI bounds are obtained directly and no assumption needs to be made on their underlying distribution (e.g. \cref{eq:PI_bounds} is not used). Furthermore, the PI quality metrics (i.e. $PICP$ and $MPIW$) may be considered during training. Such an approach is outlined by Pearce et al. \cite{pearce2018highquality} wherein a 'quality-driven' (QD) loss function incorporating proxies of the PI quality metrics are used to train an ensemble of feed-forward neural networks (FNNs). The output of each network are the upper and lower bounds of the PI associated with a prediction. The prediction itself may be reconstructed from these bounds, for example, by assuming their midpoint. The QD loss function is described in \cref{eq:QD}. 

\begin{equation}\label{eq:QD}
    QD = MPIW_{C} + N\lambda\max{(0, (1 -\alpha) - PICP_{s})}^{2}
\end{equation}

In \cref{eq:QD}, $MPIW_{C} \subseteq MPIW$ describes the $MPIW$ only of samples \textit{which encompass their measurement value} (i.e. MPIW for samples where the measurement lies outside the PI bounds are ignored). $N$ is the batch size used for training and $\lambda$ is a tunable parameter that scales the importance of the $PICP$ condition. $PICP_{s}$ represents the 'soft' $PICP$, defined in \cref{eq:PICP_s}, where $S(\cdot)$ denotes the sigmoid function and $s$ is some softening factor. 'softening' of the $PICP$ is encouraged to mitigate the non-linearity of the second term in \cref{eq:QD} for compatibility with the optimizers of existing neural network library implementations (e.g. \texttt{tensorflow}). Pearce et al. \cite{pearce2018highquality} suggest a value of $s=160$ for consistent performance in regression tasks.

\begin{equation}\label{eq:PICP_s}
    PICP_{s} = \frac{1}{N}\sum_{i=1}^{N}{S(s(y_{i} - L_{i}))\cdot S(s(U_{i} - y_{i}))}
\end{equation}


\newpage
\section{Numerical Validation of Prediction Intervals}\label{sec:numerical_validation}

In order to determine if the ANN PI estimation techniques do indeed produce valid PIs for the quadrotor, a numerical validation scheme is constructed. Equivalent polynomial models with associated PIs are also evaluated for reference.


\subsection{Simulation Model Identification}\label{subsec:system_identification}
For all modelling techniques, individual aerodynamic models are identified for the forces and moments of each of the axes (i.e. 6 models total, per identification technique). Moreover, a prediction interval criterion of 95\% is specified for each PI estimation technique (i.e. each of the estimated PIs should contain at least 95\% of the variation due to uncertainties in the data). The model identification scripts are programmed in Python (version 3.8.13). 

Polynomial aerodynamic models are identified through stepwise regression \cite{Klein2006_AircraftSysID} following the same procedure as outlined by Sun et al. \cite{Sun2018_GrayBox} for their grey-box models. The basic principle of stepwise regression is to identify the best model structure given a pool of candidate regressors. This is accomplished by sequentially adding the best-fitting regressor, orthogonalized with respect to the regressors currently in the model structure, and evaluating the current model structure for redundancy after regressor additions. In consultation with the authors of \cite{Sun2018_GrayBox}, the rotor speeds are also incorporated in the candidate regressors of $F_{z}$ for improved performance. The candidate regressors, along with the final model structures, can be found in \cref{app:Sim_Polys_candRegs} and \cref{app:Sim_Polys_modelStruct} respectively. 

The architectures of the ANN models for both PI estimation techniques are identical with only method-specific differences. Inspired by the ANNs of \cite{Bansal2016}, all ANN models used here are composed of an ensemble of 10 dense FNNs, each with an input layer, a single hidden layer with 20 neurons, and an output layer. ReLU and a linear function constitute the hidden layer and output layer activation functions respectively. ADAM is used as the optimizer. The mean-squared-error and QD cost functions are employed for the bootstrap and quality-driven methods respectively. All ANNs are trained for 50 epochs. For the bootstrap method, an additional FNN with identical structure to the base ANNs is used, with the exception of 25 hidden neurons and a modified cost function given by \cref{eq:bootstrap_cost_function}. \texttt{tensorflow} (version 2.2.0) is used to construct and train the ANNs. The input states for each ANN model can be found in \cref{app:Sim_ANN}.

\subsection{Data Generation}\label{subsec:data_generation}


\begin{figure}[!b]
    \begin{subfigure}{.48\textwidth}
        \centering
        \includegraphics[width = \textwidth]{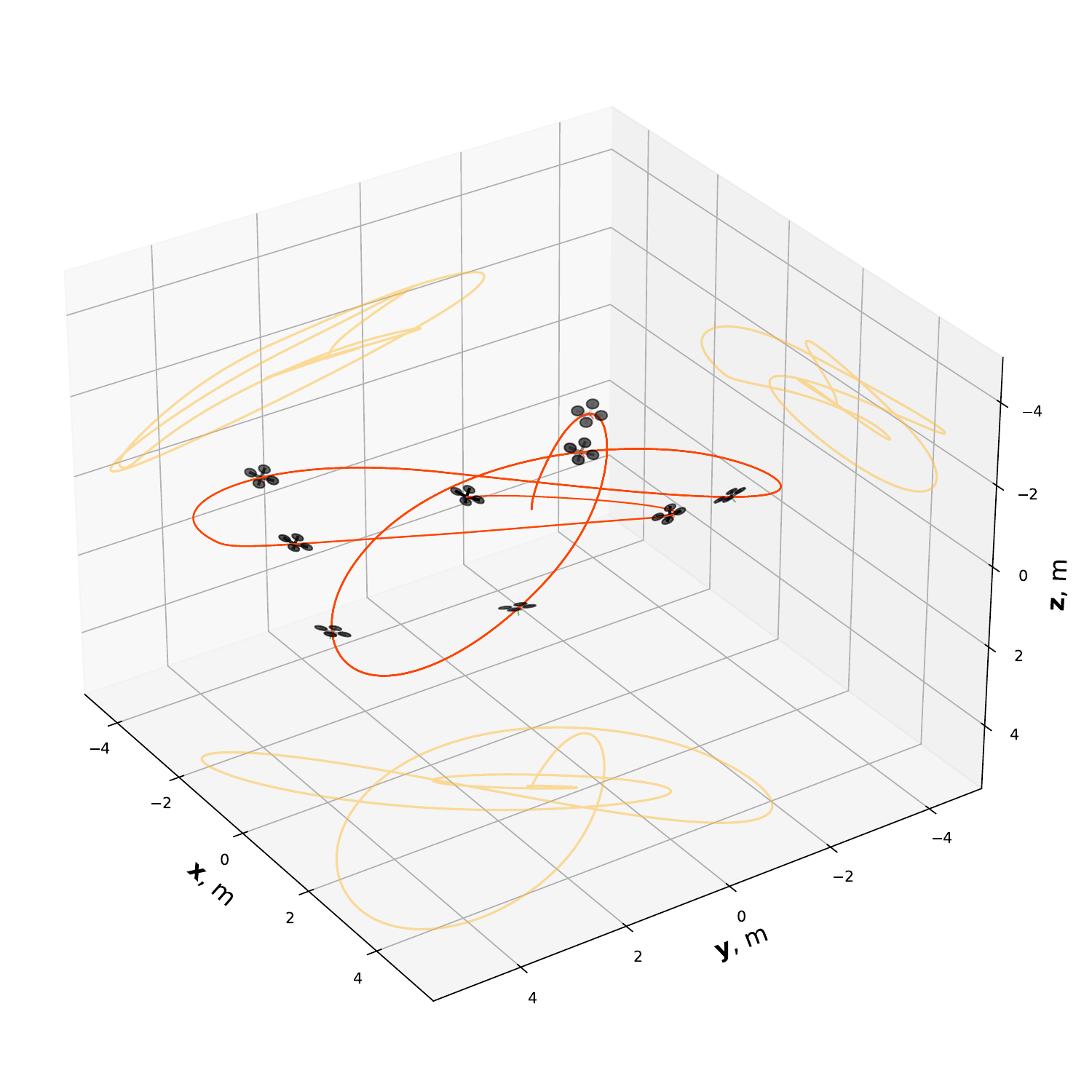}
        \caption{Sinusoidal trajectory flown in the simulation of \cite{Sun_ControlDoubleFailure} to gather mock-flight data.}
        \label{fig:simulation_trajectory}
    \end{subfigure}%
    \hfill
    \begin{subfigure}{.48\textwidth}
        \centering
        \includegraphics[width=\textwidth]{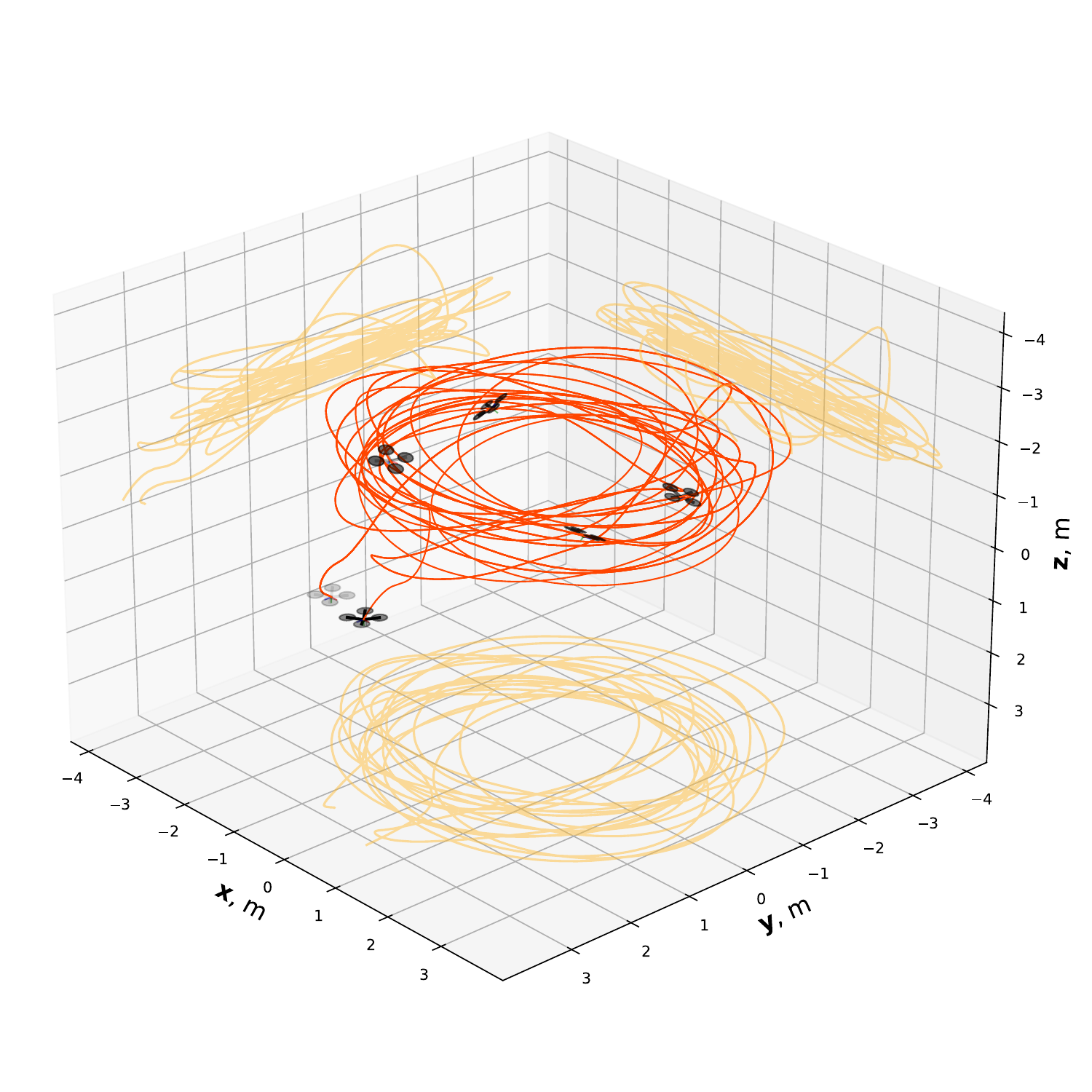}
        \caption{Circular flight trajectory flown in the CyberZoo to achieve velocities of up to 5 $ms^{-1}$.}
        \label{fig:CyberZoo_trajectory}
    \end{subfigure}%
    \caption{Example trajectories (orange) of flights employed for model identification. Also shown are the planar trajectories (yellow) of these manoeuvres, along with some snapshots of the quadrotor's orientation at various points in the trajectory.}
    \label{fig:trajectories}
\end{figure}

An existing high-fidelity quadrotor simulation, developed by Sun et al. \cite{Sun_ControlDoubleFailure} in MATLAB\textregistered (version R2022a), is used to generate mock-flight data for quadrotor model identification and PI numerical validation under consistent conditions. The simulation model itself is a rotor-local polynomial model (i.e. aerodynamic forces and moments are modelled per rotor), whereas the models identified in this paper consider only their cumulative effect on the quadrotor frame (i.e point-mass forces and moments). This discrepancy introduces some model uncertainty, which should be accountable by the estimated PIs.  

In the simulation, users can specify positional ($x_{E}$, $y_{E}$, $z_{E}$) and yaw angle ($\psi$) commands. These are used to generate a collection of simulated flight states and aerodynamic forces \& moments for model identification (training set) and numerical validation (validation set). All reference signals are either step-like pulses or sinusoidal in nature, with varying amplitude and frequency. The training set consists of both decoupled and coupled force and moment manoeuvres (e.g. decoupled: sinusoid along x-axis only; coupled: sinusoid along multiple axes). An example coupled trajectory used for training is illustrated in \cref{fig:simulation_trajectory}. The validation set only consists of coupled manoeuvres, distinct from those seen during training (i.e. different couplings, amplitudes, and sinusoid frequencies). 

As the simulation relies on a non-linear controller \cite{Sun_ControlDoubleFailure}, the flight data is initially generated noise-free to avoid the non-Gaussian transformation of input uncertainties. Noise, sampled from a Gaussian distribution, is instead injected into the relevant model inputs (i.e. \cref{eq:state_vector}) and measurements before identification and numerical validation to simulate uncertainty. The employed noise statistics and illustrative examples of these noisy realizations are depicted in \cref{app:gaussianNoiseStats}. The noise variances are chosen to be similar in magnitude to, if not worse than, widely used Inertial Measurement Unit (IMU) sensors such as the MPU6000 \cite{gonzalez2019MPU600performance}.  


For numerical validation, the identified models are tasked with making predictions on 1000, independent noise contaminated, realizations of the validation set of flight trajectories. The subsequent spread in model predictions and measurements is observed, and the ability of the PIs to contain these evaluated for each run. 

\subsection{Numerical Validation Results}\label{subsec:PI_results}

\begin{table}[!b]
\caption{Results of the numerical validation of prediction intervals (PIs), defined for a confidence level of $95\%$, subject to Gaussian noise injected into the model inputs and measurements. The containment of the variation in model predictions due to uncertain inputs is shown in the \textit{Model Variation} columns. The \textit{Measurement Variation} columns summarize the containment of measurement uncertainty. For reference, the containment of the models with respect to the combined variation for the \textit{Training} data is also shown.}
\begin{subtable}{.48\linewidth}
\centering
\caption{Force models}
\label{tab:force_results}
\resizebox{0.95\textwidth}{!}{%
\begin{tabular}{llrrrrrr}
\multicolumn{2}{c}{\multirow{2}{*}{}} &\multicolumn{1}{l}{} &\multicolumn{1}{l}{\textbf{PICP, \%}} &\multicolumn{1}{l}{} &\multicolumn{1}{l}{} &\multicolumn{1}{l}{\textbf{MPIW, \%}} &\multicolumn{1}{l}{} \\ \cline{3-8} 
  &\multicolumn{1}{c|}{} &\multirow{2}{*}{\textit{Training}} &\multicolumn{1}{c}{Model} &\multicolumn{1}{c|}{Measurement} &\multirow{2}{*}{\textit{Training}} &\multicolumn{1}{c}{Model} &\multicolumn{1}{c|}{Measurement} \\ 
  &\multicolumn{1}{c|}{} &  &\multicolumn{1}{c}{Variation} &\multicolumn{1}{c|}{Variation} &  &\multicolumn{1}{c}{Variation} &\multicolumn{1}{c|}{Variation} \\ \cline{2-8} 
  &\multicolumn{1}{|l|}{\begin{tabular}[c]{@{}l@{}}ANN\\ (\textit{Bootstrap})\end{tabular}} &\multicolumn{1}{c}{\textit{96.36}} &\multicolumn{1}{c}{100.00} &\multicolumn{1}{c|}{97.24} &\multicolumn{1}{c}{\textit{3.26}} &\multicolumn{1}{c}{6.97} &\multicolumn{1}{c|}{6.09} \\
\textbf{Fx} &\multicolumn{1}{|l|}{\cellcolor[HTML]{9dc8e3}\begin{tabular}[c]{@{}l@{}}ANN\\ (\textit{Quality driven})\end{tabular}} &\multicolumn{1}{c}{\cellcolor[HTML]{9dc8e3}\textit{97.18}} &\multicolumn{1}{c}{\cellcolor[HTML]{9dc8e3}100.00} &\multicolumn{1}{c|}{\cellcolor[HTML]{9dc8e3}97.84} &\multicolumn{1}{c}{\cellcolor[HTML]{9dc8e3}\textit{3.46}} &\multicolumn{1}{c}{\cellcolor[HTML]{9dc8e3}7.00} &\multicolumn{1}{c|}{\cellcolor[HTML]{9dc8e3}6.28} \\
  &\multicolumn{1}{|l|}{Polynomial} &\multicolumn{1}{c}{\textit{97.59}} &\multicolumn{1}{c}{100.00} &\multicolumn{1}{c|}{96.89} &\multicolumn{1}{c}{\textit{3.95}} &\multicolumn{1}{c}{6.30} &\multicolumn{1}{c|}{5.63} \\\cline{2-8}
  &\multicolumn{1}{|l|}{\begin{tabular}[c]{@{}l@{}}ANN\\ (\textit{Bootstrap})\end{tabular}} &\multicolumn{1}{c}{\textit{95.78}} &\multicolumn{1}{c}{100.00} &\multicolumn{1}{c|}{96.76} &\multicolumn{1}{c}{\textit{3.15}} &\multicolumn{1}{c}{6.16} &\multicolumn{1}{c|}{5.95} \\
\textbf{Fy} &\multicolumn{1}{|l|}{\cellcolor[HTML]{9dc8e3}\begin{tabular}[c]{@{}l@{}}ANN\\ (\textit{Quality driven})\end{tabular}} &\multicolumn{1}{c}{\cellcolor[HTML]{9dc8e3}\textit{97.54}} &\multicolumn{1}{c}{\cellcolor[HTML]{9dc8e3}100.00} &\multicolumn{1}{c|}{\cellcolor[HTML]{9dc8e3}98.33} &\multicolumn{1}{c}{\cellcolor[HTML]{9dc8e3}\textit{3.60}} &\multicolumn{1}{c}{\cellcolor[HTML]{9dc8e3}7.28} &\multicolumn{1}{c|}{\cellcolor[HTML]{9dc8e3}7.08} \\
  &\multicolumn{1}{|l|}{Polynomial} &\multicolumn{1}{c}{\textit{96.74}} &\multicolumn{1}{c}{100.00} &\multicolumn{1}{c|}{96.51} &\multicolumn{1}{c}{\textit{3.82}} &\multicolumn{1}{c}{5.67} &\multicolumn{1}{c|}{5.61} \\\cline{2-8}
  &\multicolumn{1}{|l|}{\begin{tabular}[c]{@{}l@{}}ANN\\ (\textit{Bootstrap})\end{tabular}} &\multicolumn{1}{c}{\textit{97.52}} &\multicolumn{1}{c}{99.99} &\multicolumn{1}{c|}{98.90} &\multicolumn{1}{c}{\textit{5.37}} &\multicolumn{1}{c}{7.62} &\multicolumn{1}{c|}{6.12} \\
\textbf{Fz} &\multicolumn{1}{|l|}{\cellcolor[HTML]{9dc8e3}\begin{tabular}[c]{@{}l@{}}ANN\\ (\textit{Quality driven})\end{tabular}} &\multicolumn{1}{c}{\cellcolor[HTML]{9dc8e3}\textit{97.59}} &\multicolumn{1}{c}{\cellcolor[HTML]{9dc8e3}100.00} &\multicolumn{1}{c|}{\cellcolor[HTML]{9dc8e3}99.40} &\multicolumn{1}{c}{\cellcolor[HTML]{9dc8e3}\textit{3.02}} &\multicolumn{1}{c}{\cellcolor[HTML]{9dc8e3}9.25} &\multicolumn{1}{c|}{\cellcolor[HTML]{9dc8e3}6.92} \\
  &\multicolumn{1}{|l|}{Polynomial} &\multicolumn{1}{c}{\textit{96.68}} &\multicolumn{1}{c}{100.00} &\multicolumn{1}{c|}{99.94} &\multicolumn{1}{c}{\textit{8.13}} &\multicolumn{1}{c}{27.32} &\multicolumn{1}{c|}{21.78} \\\cline{2-8}
\end{tabular}%
}
\end{subtable}%
\begin{subtable}{.48\linewidth}
\centering
\caption{Moment models}
\label{tab:moment_results}
\resizebox{0.95\textwidth}{!}{%
\begin{tabular}{llrrrrrr}
\multicolumn{2}{c}{\multirow{2}{*}{}} &\multicolumn{1}{l}{} &\multicolumn{1}{l}{\textbf{PICP, \%}} &\multicolumn{1}{l}{} &\multicolumn{1}{l}{} &\multicolumn{1}{l}{\textbf{MPIW, \%}} &\multicolumn{1}{l}{} \\ \cline{3-8} 
  &\multicolumn{1}{c|}{} &\multirow{2}{*}{\textit{Training}} &\multicolumn{1}{c}{Model} &\multicolumn{1}{c|}{Measurement} &\multirow{2}{*}{\textit{Training}} &\multicolumn{1}{c}{Model} &\multicolumn{1}{c|}{Measurement} \\ 
  &\multicolumn{1}{c|}{} &  &\multicolumn{1}{c}{Variation} &\multicolumn{1}{c|}{Variation} &  &\multicolumn{1}{c}{Variation} &\multicolumn{1}{c|}{Variation} \\ \cline{2-8} 
  &\multicolumn{1}{|l|}{\begin{tabular}[c]{@{}l@{}}ANN\\ (\textit{Bootstrap})\end{tabular}} &\multicolumn{1}{c}{\textit{96.05}} &\multicolumn{1}{c}{98.63} &\multicolumn{1}{c|}{97.67} &\multicolumn{1}{c}{\textit{2.40}} &\multicolumn{1}{c}{3.83} &\multicolumn{1}{c|}{3.84} \\
\textbf{Mx} &\multicolumn{1}{|l|}{\cellcolor[HTML]{9dc8e3}\begin{tabular}[c]{@{}l@{}}ANN\\ (\textit{Quality driven})\end{tabular}} &\multicolumn{1}{c}{\cellcolor[HTML]{9dc8e3}\textit{98.51}} &\multicolumn{1}{c}{\cellcolor[HTML]{9dc8e3}99.55} &\multicolumn{1}{c|}{\cellcolor[HTML]{9dc8e3}98.40} &\multicolumn{1}{c}{\cellcolor[HTML]{9dc8e3}\textit{2.96}} &\multicolumn{1}{c}{\cellcolor[HTML]{9dc8e3}3.87} &\multicolumn{1}{c|}{\cellcolor[HTML]{9dc8e3}3.81} \\
  &\multicolumn{1}{|l|}{Polynomial} &\multicolumn{1}{c}{\textit{96.22}} &\multicolumn{1}{c}{98.75} &\multicolumn{1}{c|}{98.29} &\multicolumn{1}{c}{\textit{3.14}} &\multicolumn{1}{c}{5.58} &\multicolumn{1}{c|}{5.80} \\\cline{2-8}
  &\multicolumn{1}{|l|}{\begin{tabular}[c]{@{}l@{}}ANN\\ (\textit{Bootstrap})\end{tabular}} &\multicolumn{1}{c}{\textit{96.84}} &\multicolumn{1}{c}{98.26} &\multicolumn{1}{c|}{97.89} &\multicolumn{1}{c}{\textit{2.76}} &\multicolumn{1}{c}{5.76} &\multicolumn{1}{c|}{6.39} \\
\textbf{My} &\multicolumn{1}{|l|}{\cellcolor[HTML]{9dc8e3}\begin{tabular}[c]{@{}l@{}}ANN\\ (\textit{Quality driven})\end{tabular}} &\multicolumn{1}{c}{\cellcolor[HTML]{9dc8e3}\textit{99.13}} &\multicolumn{1}{c}{\cellcolor[HTML]{9dc8e3}99.65} &\multicolumn{1}{c|}{\cellcolor[HTML]{9dc8e3}98.84} &\multicolumn{1}{c}{\cellcolor[HTML]{9dc8e3}\textit{3.75}} &\multicolumn{1}{c}{\cellcolor[HTML]{9dc8e3}6.11} &\multicolumn{1}{c|}{\cellcolor[HTML]{9dc8e3}7.20} \\
  &\multicolumn{1}{|l|}{Polynomial} &\multicolumn{1}{c}{\textit{96.04}} &\multicolumn{1}{c}{98.54} &\multicolumn{1}{c|}{98.00} &\multicolumn{1}{c}{\textit{2.95}} &\multicolumn{1}{c}{7.19} &\multicolumn{1}{c|}{8.18} \\\cline{2-8}
  &\multicolumn{1}{|l|}{\begin{tabular}[c]{@{}l@{}}ANN\\ (\textit{Bootstrap})\end{tabular}} &\multicolumn{1}{c}{\textit{98.05}} &\multicolumn{1}{c}{100.00} &\multicolumn{1}{c|}{98.85} &\multicolumn{1}{c}{\textit{6.36}} &\multicolumn{1}{c}{8.77} &\multicolumn{1}{c|}{8.41} \\
\textbf{Mz} &\multicolumn{1}{|l|}{\cellcolor[HTML]{9dc8e3}\begin{tabular}[c]{@{}l@{}}ANN\\ (\textit{Quality driven})\end{tabular}} &\multicolumn{1}{c}{\cellcolor[HTML]{9dc8e3}\textit{97.86}} &\multicolumn{1}{c}{\cellcolor[HTML]{9dc8e3}100.00} &\multicolumn{1}{c|}{\cellcolor[HTML]{9dc8e3}97.89} &\multicolumn{1}{c}{\cellcolor[HTML]{9dc8e3}\textit{5.28}} &\multicolumn{1}{c}{\cellcolor[HTML]{9dc8e3}7.08} &\multicolumn{1}{c|}{\cellcolor[HTML]{9dc8e3}6.57} \\
  &\multicolumn{1}{|l|}{Polynomial} &\multicolumn{1}{c}{\textit{96.75}} &\multicolumn{1}{c}{100.00} &\multicolumn{1}{c|}{96.92} &\multicolumn{1}{c}{\textit{4.64}} &\multicolumn{1}{c}{6.00} &\multicolumn{1}{c|}{5.77} \\\cline{2-8}
\end{tabular}%
}
\end{subtable}
\end{table}

\begin{figure}[!t]
    \centering
    \includegraphics[width = 0.98\textwidth]{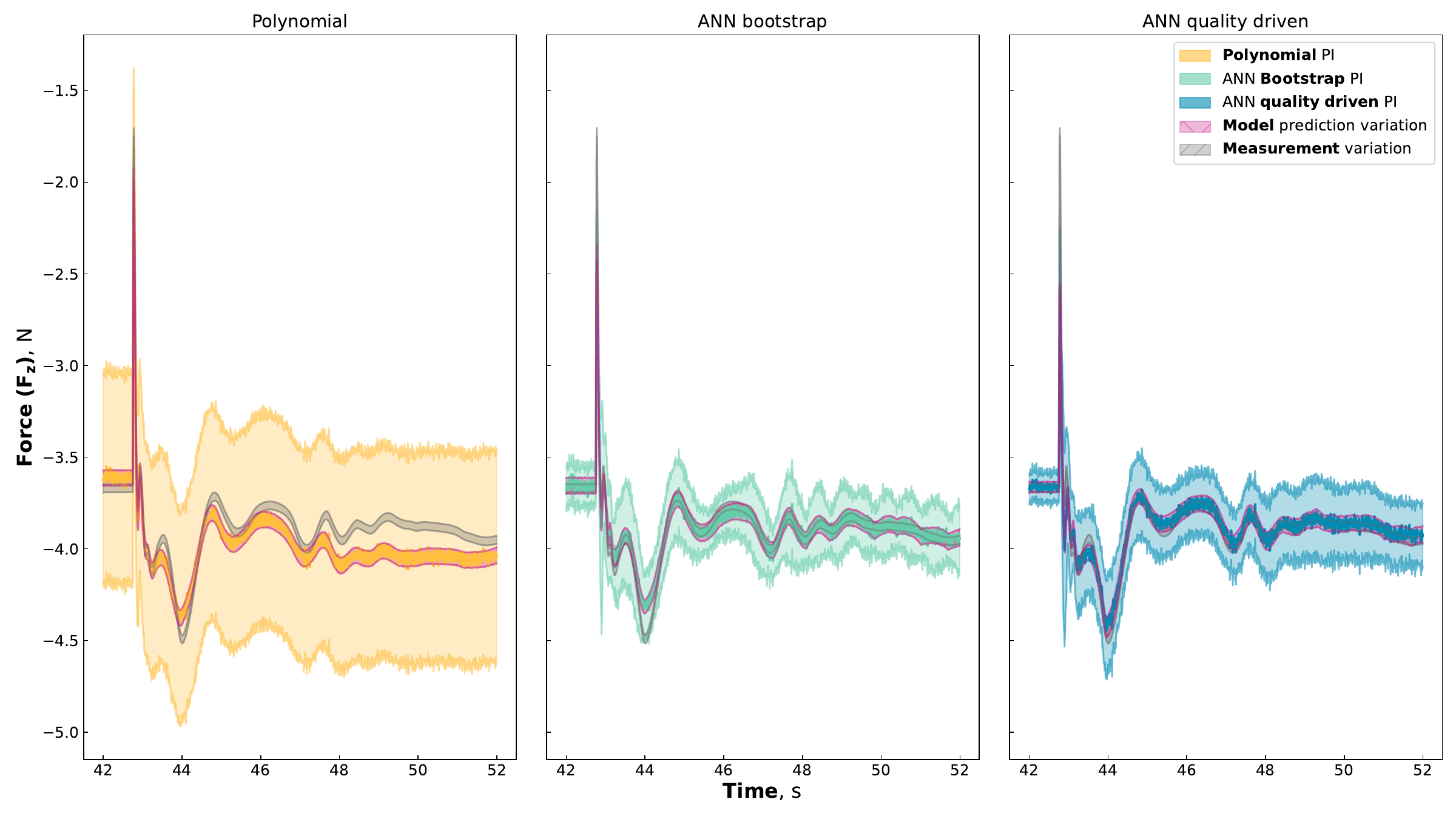}
    \caption{Illustration of the 95\% prediction interval (PI) performances of each of the PI estimation techniques - Polynomial (orange), ANN bootstrap (Cyan), ANN quality driven (Blue) - with respect to Gaussian uncertainty in the $F_{z}$ measurements (grey) and model inputs, which culminate in model prediction variation (pink).}
    \label{fig:Fz_PI_Gaussian_BothNoise_SinusoidalOnly}
\end{figure}

\Cref{tab:force_results} and \cref{tab:moment_results} summarize the results of the numerical validation of the estimated PIs subject to Gaussain noise contamination. By definition, PIs can only be considered valid when both the variation in model predictions due to uncertain inputs and uncertain measurements satisfy the PICP condition (here: $\ge 95.00\%$). This is distinguished in the tables through the \textit{Model Variation} and \textit{Measurement Variation} columns respectively. For reference, the PICP and (normalized) MPIW with respect to the total variation in the training data are also shown through the \textit{Training} column. 

Indeed, \cref{tab:force_results} and \cref{tab:moment_results} show that all estimation techniques comfortably encompass at least 95\% of the variations due to both uncertain inputs and uncertain measurements. Therefore, all considered estimation methods produce valid PIs. While this is perhaps unsurprising for the polynomial models, this is a significant result for the ANN PI estimation techniques as it demonstrates their capacity to estimate valid PIs for quadrotor aerodynamic models subject to Gaussian noise contamination.

\begin{figure}[!t]
    \centering
    \includegraphics[width = 0.98\textwidth]{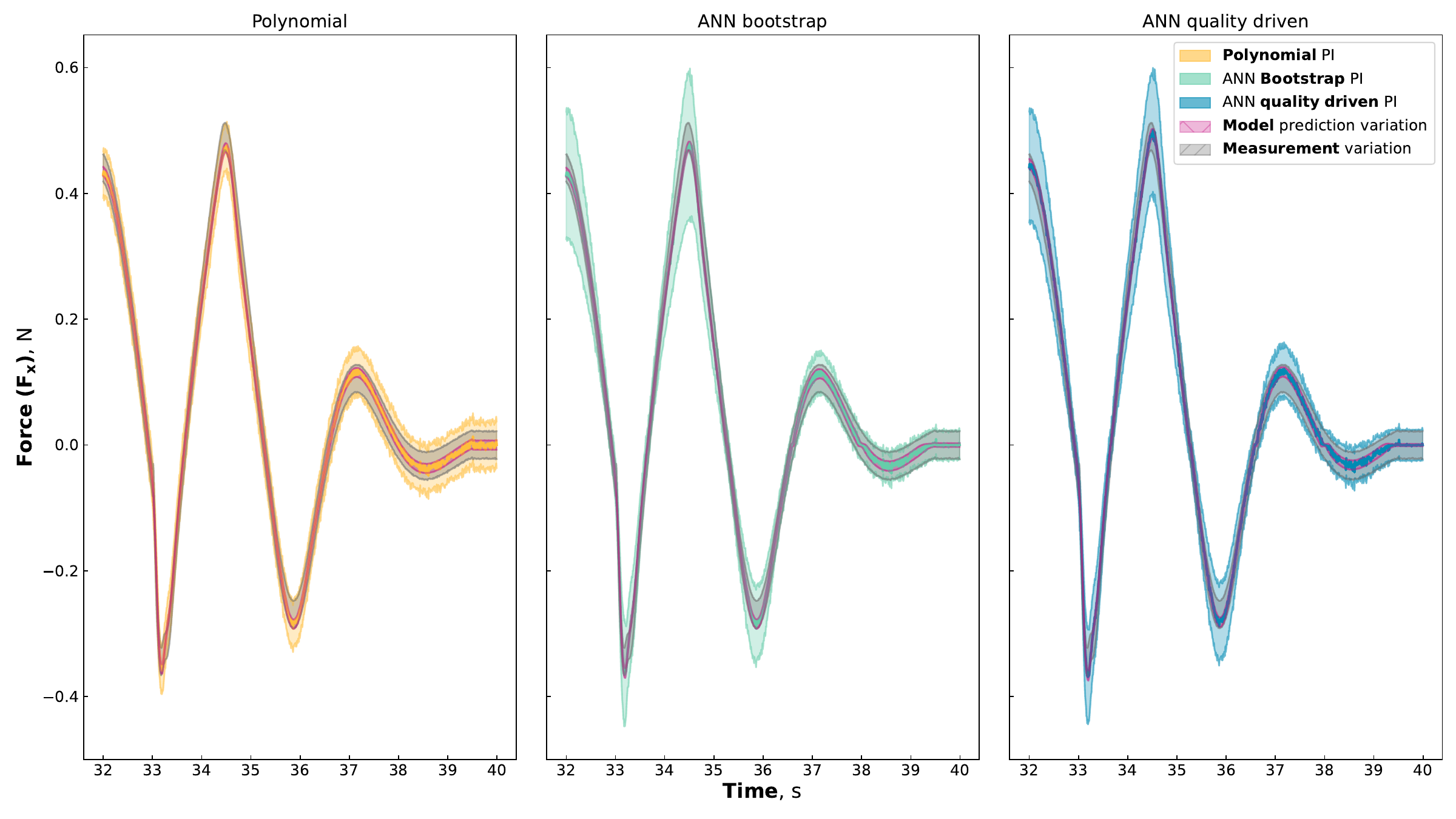}
    \caption{Illustration of the 95\% prediction interval (PI) performances of each of the PI estimation techniques - Polynomial (orange), ANN bootstrap (Cyan), ANN quality driven (Blue) - with respect to Gaussian uncertainty in the $F_{x}$ measurements (grey) and model inputs, which culminate in model prediction variation (pink).}
    \label{fig:Fx_PI_Gaussian_BothNoise_SinusoidalOnly}
\end{figure}

The ANN quality driven method produces the widest PIs on average (compare the $MPIW$ of the methods in \cref{tab:force_results} and \cref{tab:moment_results}). This result may stem from the fact that the ANN quality driven method requires more training epochs than the bootstrap method to converge \cite{KhosraviLUBE,pearce2018highquality} and, here, they are trained for the same amount of epochs. Hence, the PI performance of the quality driven method may improve if afforded more training epochs, at the risk of over-fitting. The ANN bootstrap and polynomial methods produce PIs of similar widths, with the exception of $F_{z}$ wherein the polynomial model estimates excessively wide PIs. This poor performance is a symptom of the lack of approximation power of the polynomial $F_{z}$ model which, at times, struggles to fit the measurement data. This is evident in the under-approximation\footnote{Naturally, this under-approximation calls into question the validity of the polynomial $F_{z}$ model with respect to the Ordinary Least Squares (OLS) assumption of zero-mean error residuals. However, this assumption is satisfied for this model with $-3.17\cdot10^{-14} \approx 0$.} of this model in \cref{fig:Fz_PI_Gaussian_BothNoise_SinusoidalOnly} and both the number of selected regressors and relatively poor coefficient of determination (R2) of this model in \cref{tab:simModel_ForceStructures}. The poor $F_{z}$ performance likely occurs because the simulation model of Sun et al. \cite{Sun_ControlDoubleFailure} is rotor-local (i.e. aerodynamic models for each rotor) whereas the polynomial represents their aggregated effect. To account for this discrepancy and maintain valid PIs, the error residuals in \cref{eq:sigma_est} must grow.

\begin{figure}[!t]
    \centering
    \includegraphics[width = 0.97\textwidth]{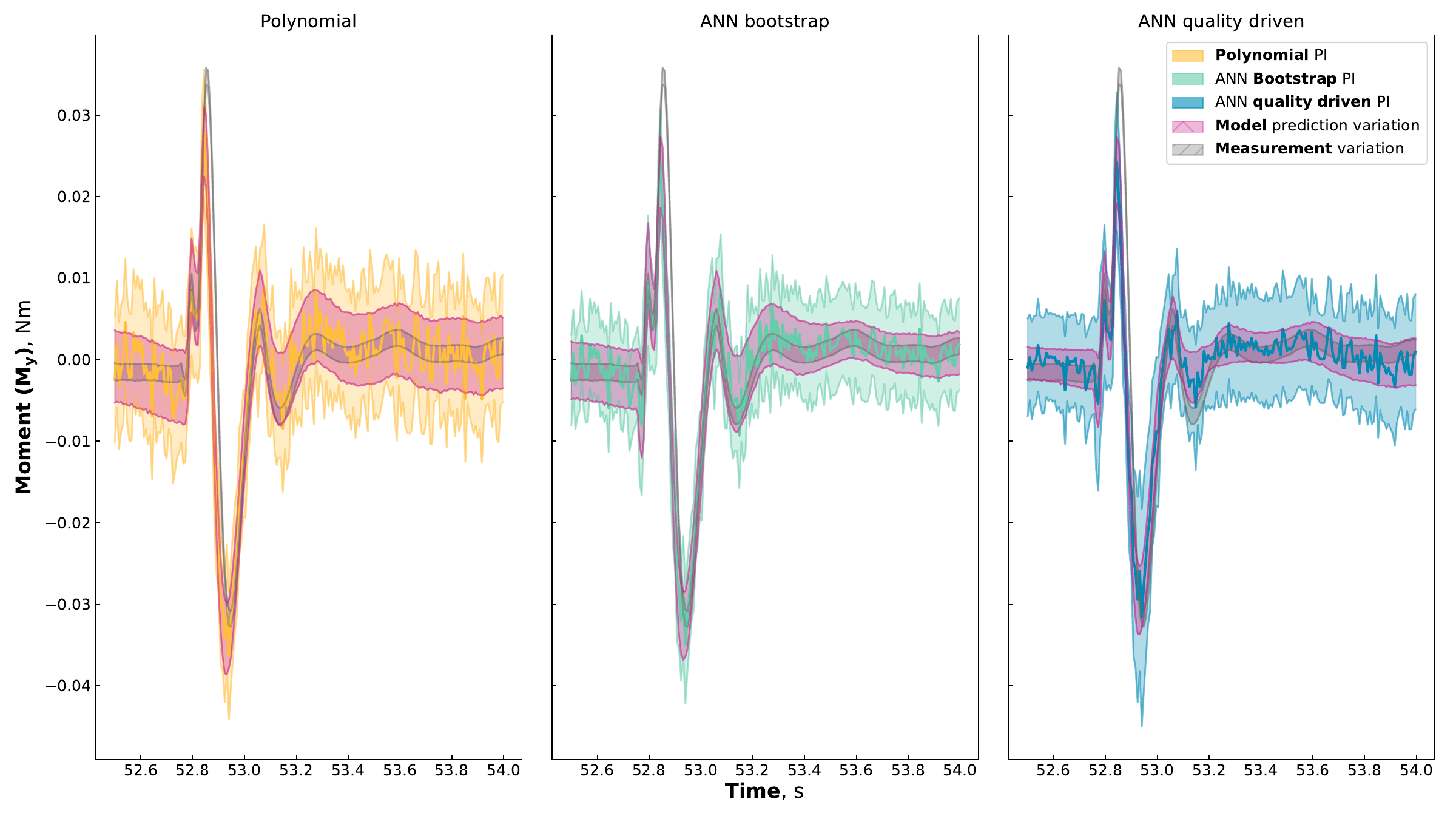}
    \caption{Illustration of the 95\% prediction interval (PI) performances of each of the PI estimation techniques - Polynomial (orange), ANN bootstrap (Cyan), ANN quality driven (Blue) - with respect to Gaussian uncertainty in the $M_{y}$ measurements (grey) and model inputs, which culminate in model prediction variation (pink).}
    \label{fig:My_PI_Gaussian_BothNoise_SinusoidalOnly}
\end{figure}

As illustrative examples of when all model identification techniques enjoy sufficient approximation power,
\cref{fig:Fx_PI_Gaussian_BothNoise_SinusoidalOnly} and \cref{fig:My_PI_Gaussian_BothNoise_SinusoidalOnly} compare the estimated PIs for $F_{x}$ and $M_{y}$ respectively. Interestingly, although the PI performances are comparable between all techniques (see \cref{tab:force_results} and \cref{tab:moment_results}), their behaviour is visually different. The polynomial PIs remain relatively consistent in width across manoeuvres and trimmed flight whereas the ANN-based PIs grow and contract with the magnitude of the forces and moments, reflecting uncertainty in the model. This an important attribute of the these PIs that will be revisited as their model extrapolation sensitivity is explored with real quadrotor flight data in \cref{sec:application_real_quad_data}. 

Overall, these numerical validation results suggest that the bootstrap method is the superior ANN PI estimation technique. Moreover, the fact that this method is compatible with already trained ensembles of ANNs only further promotes its utility. Note, however, that validity of the PIs is only demonstrated for Gaussian noise and uncertainties. Furthermore, when afforded sufficient approximation power, the polynomial estimation produces the highest quality PIs.




\section{Application to Real Quadrotor Flight Data}\label{sec:application_real_quad_data}

While the numerical validation results show that the discussed PI estimation methods produce valid PIs, it is as of yet unclear how practical such PIs are when applied to real quadrotor flight data. Moreover, it would be beneficial if the estimated PIs widen when tasked with making predictions beyond the identification flight envelope but remain the same, or even decrease, when interpolating. To evaluate these, models and associated PIs are estimated on real quadrotor flight data obtained in a wind tunnel at different flight speeds.

\subsection{Quadrotor platform}\label{subsec:quadPlatform}
The \textit{HDBeetle} - a custom 3-inch quadrotor constructed from commercial components - is used as the test platform and is depicted in \cref{fig:HDBeetle}. The physical properties of the \textit{HDBeetle} are summarized in \cref{tab:HDBeetle_properities}. This quadrotor is equipped with the Holybro Kakute F7 (STM34F745) flight controller. BetaFlight is elected for the flight control software since it is open-source and supports Bi-directional DSHOT. The latter of which facilitates the high-frequency and sufficiently synchronized reading of the motor (e)RPMs\footnote{Note that the rotor RPMs are not measured directly, but rather, a proxy of this through their electronic RPMs (i.e. eRPMs). The true RPMs may be reconstructed by scaling the eRPM value by some constant, typically, the number of motor poles (= 12 for the \textit{HDBeetle} motors).}. Such information is necessary to identify accurate models of the quadrotor. A disadvantage of BetaFlight is that it has limited autonomous support, hence all flights are piloted. Subsequently, the \textit{HDBeetle} is outfitted with the DJI Digital FPV system to facilitate piloted first-person-view (FPV) flight. 

The body accelerations and rotational rates are measured by the on-board IMU (MPU6000). Moreover, BetaFlight also estimates the quadrotor's attitude through sensor fusion of the IMU data. While these states are logged at 500 Hz by BetaFlight, they are pre-filtered (through a series of notch and low-pass filters) before logging. Fortunately, the motor rotational speeds are available, also at 500 Hz, without any pre-processing.

\begin{table}[!h]
\centering
    \caption{Physical properties of the \textit{HDBeetle}. Note that mass of the quadrotor is an average, and may change slightly across flights due to battery exchanges. Moreover, the moment of inertia is obtained experimentally through a trifilar pendulum with the propellers removed.}
    \label{tab:HDBeetle_properities}
    \begin{tabular}{cccccc} \hline \hline
    \textbf{Mass}, $kg$ & \multicolumn{3}{l}{\textbf{Moment of Inertia}, $kg\cdot m^{2}$} & \textbf{Characteristic length} $b$, $m$ & \textbf{Rotor radius} $R$, $m$ \\
                                                                   & $I_{xx}$            & $I_{yy}$            & $I_{zz}$              & & \\ \hline 
    0.433 &   8.65E-04	& 1.07E-03 &	1.71E-03   & 0.077 & 0.0381 \\ \hline \hline
    \end{tabular}
\end{table}


\subsection{Data Acquisition}\label{subsec:realDataAcq}

\begin{figure}[!h]
    \begin{minipage}{.48\textwidth}
        \centering
        \includegraphics[width=\textwidth]{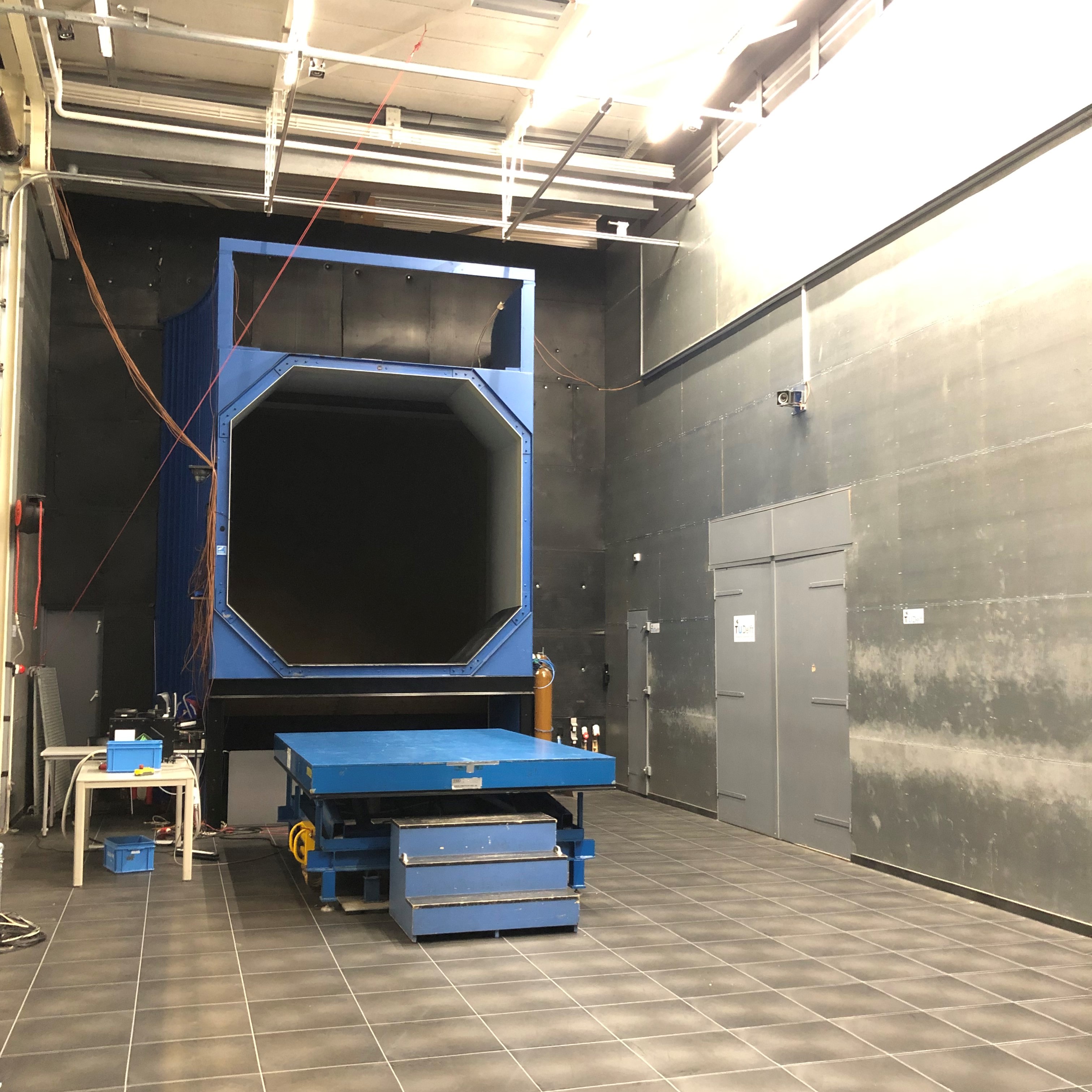}
        \caption{The Open Jet Facility (OJF) of the Aerospace Engineering faculty at the Delft University of Technology. The OJF has a test section (i.e. nozzle) of 2.85 by 2.85 $m$ and is capable of achieving wind-speeds of up to 35 $ms^{-1}$.}
        \label{fig:OJF}
    \end{minipage}%
    \hfill
    \begin{minipage}{.48\textwidth}
        \centering
        \includegraphics[width=\textwidth]{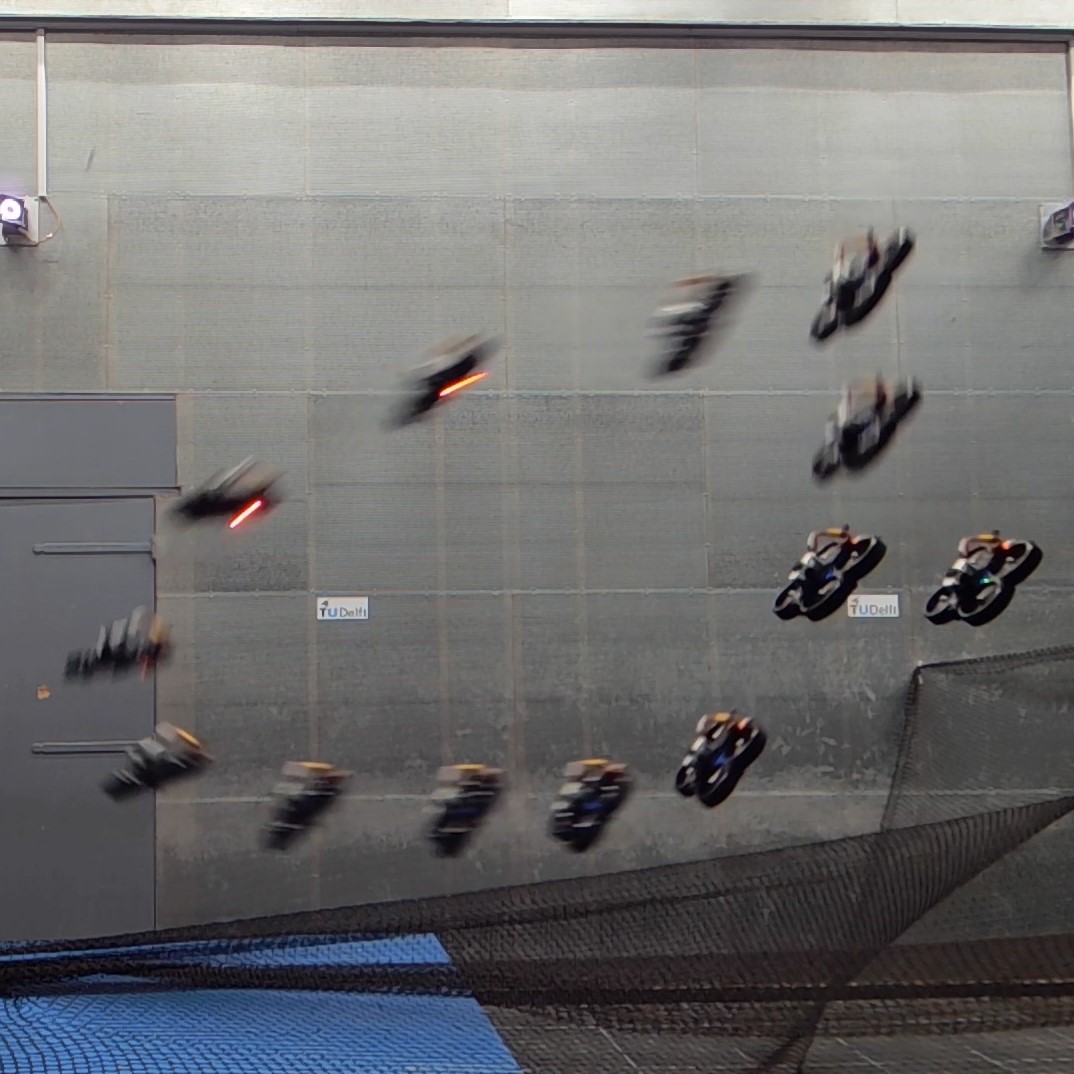}
        \caption{Example trajectory of a forward-backward manoeuvre to excite the $F_{x}$ and $F_{z}$ forces during a 10 $ms^{-1}$ flight in the Open Jet Facility. The nozzle of the wind tunnel is to the left of the picture, thus the incident wind moves from left to right.}
        \label{fig:OJF_Excitations}
    \end{minipage}%
\end{figure}

The quadrotor is flown in both the CyberZoo and Open Jet Facility (\cref{fig:OJF}) of the Faculty of Aerospace Engineering at the Delft University of Technology. Both facilities are equipped with the OptiTrack motion capturing system, which completes the state vector by providing accurate attitude and velocity information at 120 Hz\footnote{To facilitate smooth tracking, a trade-off has to be made between the OptiTrack exposure and sampling rate settings. For our applications, a sampling rate of 120 Hz is the highest found that affords suitable tracking performance.}. 

The CyberZoo is an open flight arena of approximately 7x10x10 $m$ in height, width, and length respectively. This confined space only allows for hovering and low-speed flights (up to 5 $ms^{-1}$). The aim of the low-speed flights is to fly as fast as possible by orbiting around an object stationed at the center of the CyberZoo, as depicted in \cref{fig:CyberZoo_trajectory}. High-speed flights (up to 14 $ms^{-1}$) are flown in the Open Jet Facility (OJF) wherein both 'steady state' and 'active' flights are flown. The steady state flights involve keeping the quadrotor in approximately\footnote{Note that due to the piloted flights, the quadrotor is not held exactly in the same spot. Thus, the objective is to keep it within the clean airflow and as stationary as possible.} the same position while the active flights aim to excite the longitudinal forces and moments. These therefore involve flying into and away from the incident wind to excite $F_{x}$ \& $F_{z}$ (refer to \cref{fig:OJF_Excitations}) and rapidly pulsing the pitch to excite $M_{y}$. As only these forces and moments can be excited safely for piloted FPV flight in the wind-tunnel, the demonstration of PI practicality for real-world models is limited to the longitudinal forces and moments. Nonetheless, the results found should extend to the lateral models by virtue of the symmetry of the quadrotor. Both the steady state and active flights were performed at wind speeds of 5, 8, 10, and 14 $ms^{-1}$. These wind speeds are recorded through LabView at 500 Hz.

\subsection{Data Processing}\label{subsec:realDataProcessing}

Due to a lack of reliable timestamp information in the equipped version of BetaFlight, the onboard data is aligned with the OptiTrack data (re-sampled to 500 Hz) through a cross-correlation of the attitude angles from each data source for each flight. For the OJF data, the wind tunnel measurements are synchronized and added to the (re-sampled) OptiTrack velocities using timestamp information prior to the onboard alignment. 

The synchronized data is then passed through an extended Kalman Filter (EKF), based on \cite{Armanini2017}, to fuse the onboard sensor information with the (wind-combined) OptiTrack data and to estimate IMU biases. While OptiTrack is often considered ground-truth, occasional dropouts in tracking occur, for example, during high-speed flight where the quadrotor is pitched significantly and the tracking markers are thus obscured (see, for example, attitudes reached during some manoeuvres: \cref{fig:OJF_Excitations}). This results in erroneous oscillations in the OptiTrack data, which is mitigated by the IMU data in the EKF. Subsequently, the data is further filtered using a low-pass 4th order butter-worth filter at 100 Hz \cite{Sun2018_GrayBox}.  

\subsection{Model Identification}\label{subsec:realDataSysID}

\begin{figure}[H]
    \centering
    \includegraphics[width = \textwidth]{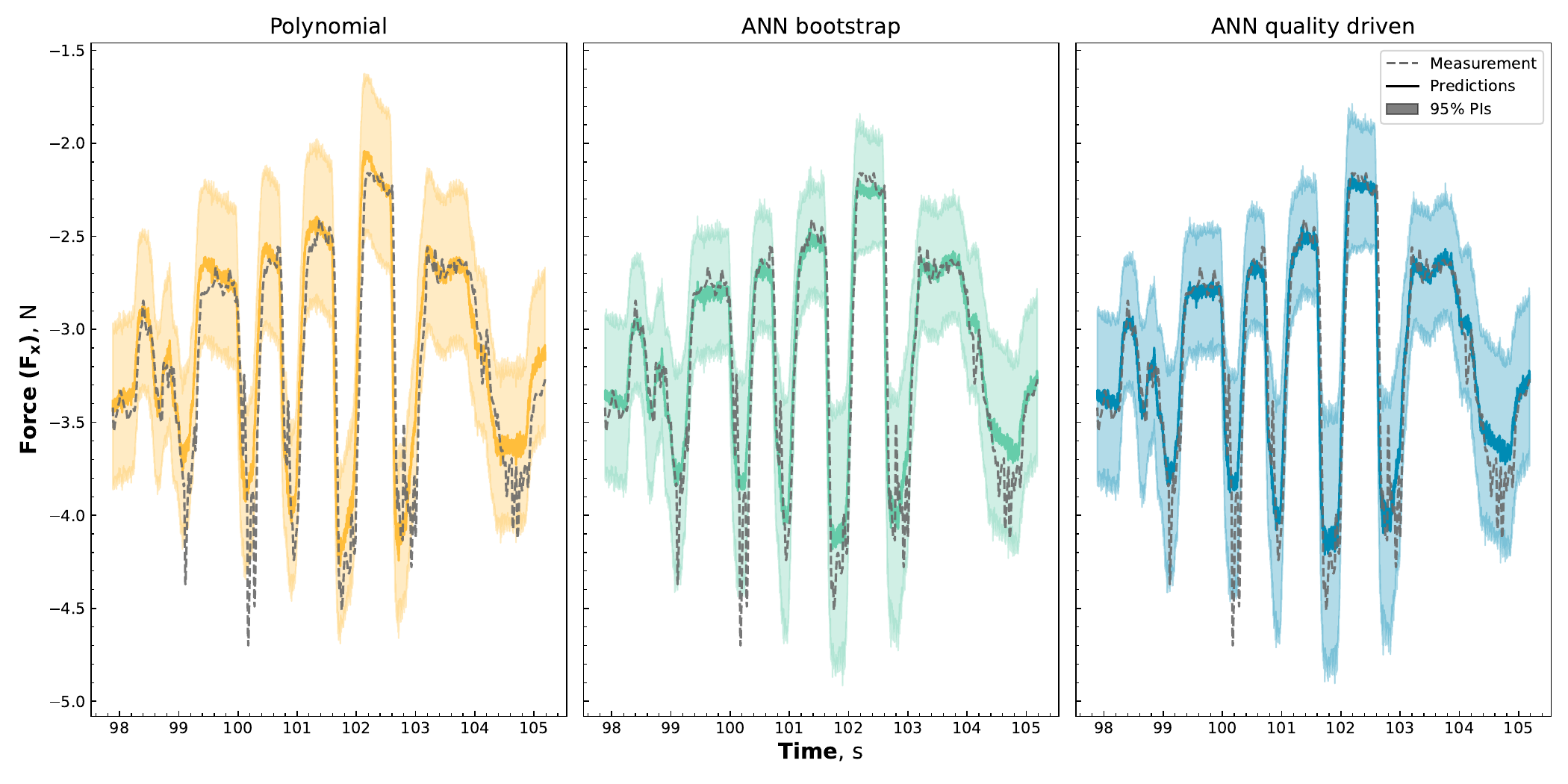}
    \caption{Model predictive capacities, and associated 95\% prediction intervals (PIs), of the identified $F_{x}$ models of the \textit{HDBeetle} during throttle pulse manoeuvres in the Open Jet Facility (wind speed = $10$ $ms^{-1})$. The dotted grey line denotes the measured $F_{x}$ and the solid lines describe the model-specific predictions. 95\% PIs are represented by the corresponding shaded region.}
    \label{fig:HDBeetle_Model_Fx}
\end{figure}

\begin{figure}[H]
    \centering
    \includegraphics[width = \textwidth]{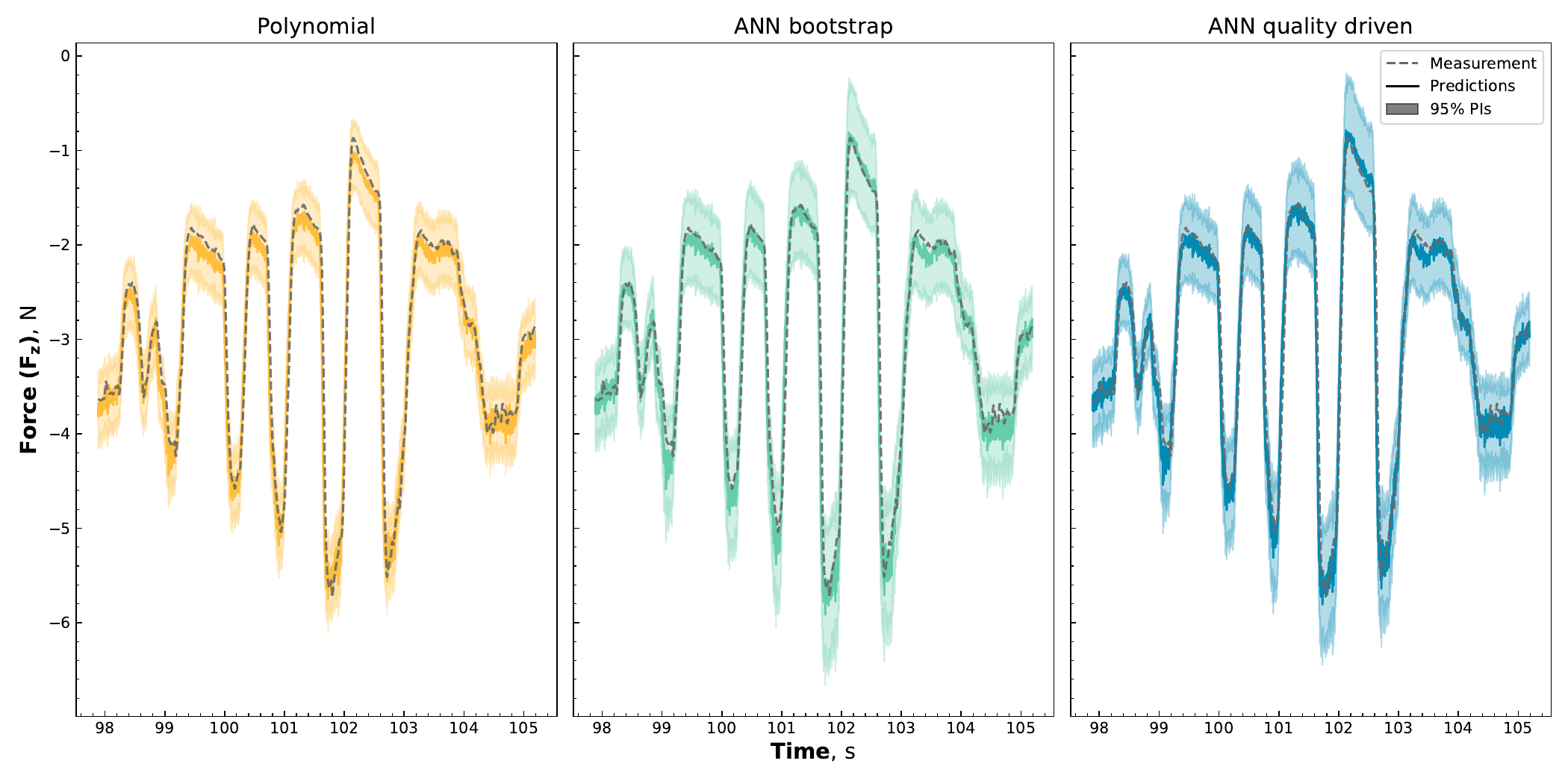}
    \caption{Model predictive capacities, and associated 95\% prediction intervals, of the identified $F_{z}$ models of the \textit{HDBeetle} during throttle pulse manoeuvres in the Open Jet Facility (wind speed = $10$ $ms^{-1})$. The dotted grey line denotes the measured $F_{z}$ and the solid lines describe the model-specific predictions. 95\% PIs are represented by the corresponding shaded region.}
    \label{fig:HDBeetle_Model_Fz}
\end{figure}

\begin{figure}[H]
    \centering
    \includegraphics[width = \textwidth]{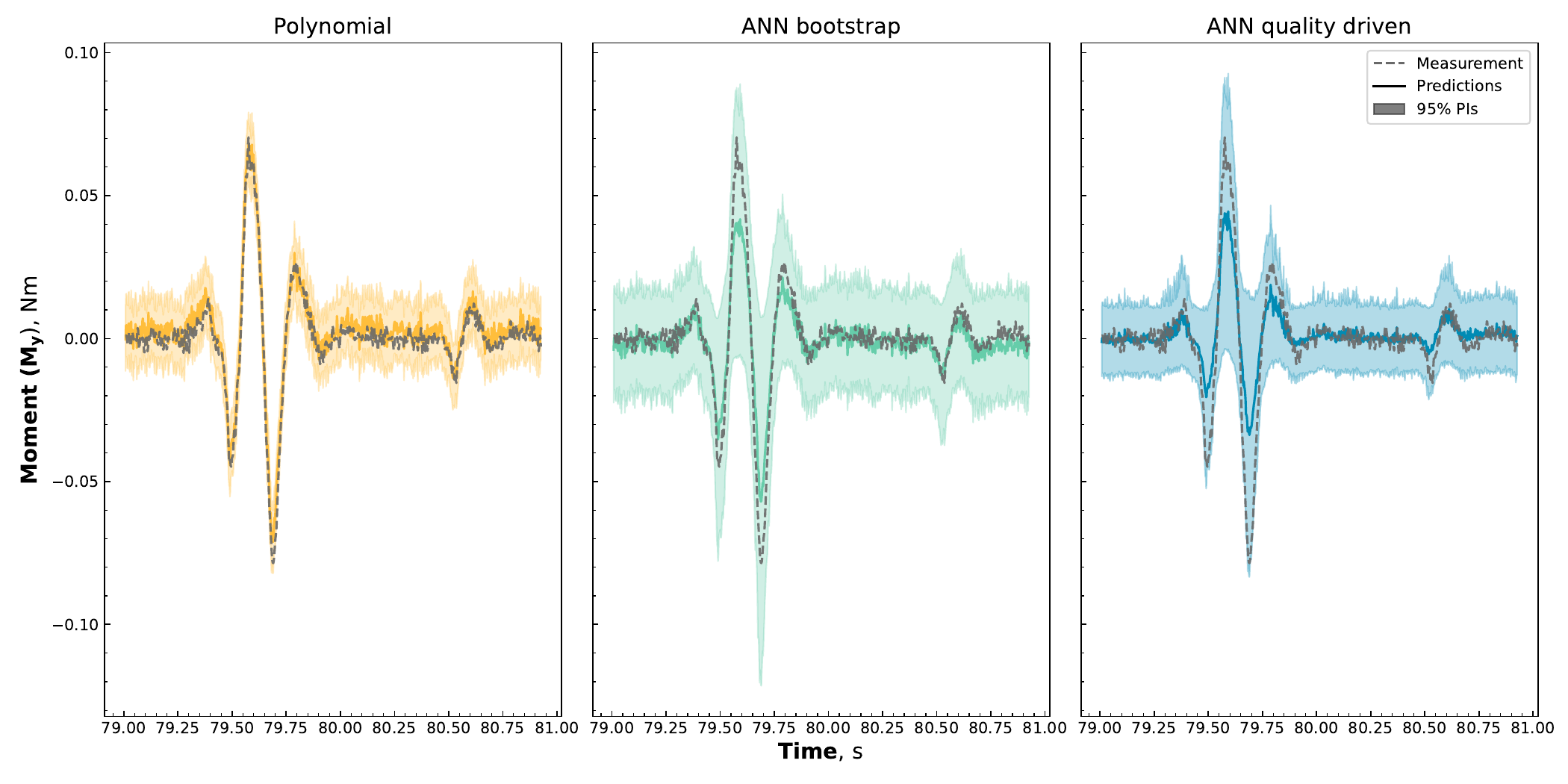}
    \caption{Model predictive capacities, and associated 95\% prediction intervals, of the identified $M_{y}$ models of the \textit{HDBeetle} during pitch angle pulse manoeuvres in the Open Jet Facility (wind speed = $10$ $ms^{-1})$. The dotted grey line denotes the measured $M_{y}$ and the solid lines describe the model-specific predictions. 95\% PIs are represented by the corresponding shaded region.}
    \label{fig:HDBeetle_Model_My}
\end{figure}

Quadrotor longitudinal aerodynamic models are then identified using the same procedure as for the simulation models in \cref{subsec:system_identification} but with candidate polynomials and ANN input vectors as shown in \cref{app:realPolys_candidateRegressors} and \cref{app:sub_Real_ANN} respectively. For identification, the body rotational rates and rotor speeds are taken from the onboard data while the velocities and attitudes stem from OptiTrack. \Cref{tab:HDBeetleModel_MomentStructures} summarizes the identified polynomial model structures. \Cref{fig:HDBeetle_Model_Fx}, \cref{fig:HDBeetle_Model_Fz}, and \cref{fig:HDBeetle_Model_My} depict the predictive capacities, and associated PIs, of the identified models of the \textit{HDBeetle} constrained to regions of relevant excitation manoeuvres. All models appear to fit the measurement data well, especially for low frequencies.

To investigate the behaviour of the estimated PIs when subjected to interpolation and extrapolation, the quadrotor models are identified on data limited to 10 $ms^{-1}$ but excluding $8$ $ms^{-1}$ wind tunnel flights. These models are then tasked with making predictions on the 0, 5, and 10 $ms^{-1}$ training (identification) data, on wind tunnel $8$ $ms^{-1}$ data (model interpolation), and on wind tunnel 14 $ms^{-1}$ data (model extrapolation). It is expected that the PIs stay approximately the same, or shrink, when interpolating but widen (through a larger MPIW) once the models begin extrapolating.

\subsection{Real-world PI Performance}\label{subsec:real_results}

\Cref{tab:realPIPerf_InAndOutOfFlightEnvelope} presents the PI performances of the longitudinal \textit{HDBeetle} aerodynamic models subject to the identification (training) data, model interpolation (8 $ms^{-1}$ data), and model extrapolation (14 $ms^{-1}$ data). To facilitate a comparison between the difference flight speeds, the absolute MPIW values are shown in \cref{tab:realPIPerf_InAndOutOfFlightEnvelope}.

\begin{table}[!b]
    \centering
    \caption{Comparison of the 95\% prediction interval (PI) performances for the identified longitudinal models of the \textit{HDBeetle} when applied to prediction on the identification (incident wind of 10 $ms^{-1}$), interpolation (incident wind of 8 $ms^{-1}$), and extrapolation (incident wind of 14 $ms^{-1}$) data sets.}
    \label{tab:realPIPerf_InAndOutOfFlightEnvelope}    
    \resizebox{0.75\textwidth}{!}{%
    \begin{tabular}{llrrrrrr}
    \multicolumn{2}{c}{} &\multicolumn{3}{c}{\textbf{PICP, \%}} &\multicolumn{3}{c}{\textbf{MPIW, N}} \\ \cline{3-8} 
      &  &\multicolumn{1}{|r}{\textit{Identification}} &\multicolumn{1}{r}{Interpolation} &\multicolumn{1}{r}{Extrapolation} &\multicolumn{1}{|r}{\textit{Identification}} &\multicolumn{1}{r}{Interpolation} &\multicolumn{1}{r|}{Extrapolation} \\ \cline{2-8} 
      &\multicolumn{1}{|l|}{\begin{tabular}[c]{@{}l@{}}ANN\\ (Bootstrap)\end{tabular}} &\multicolumn{1}{c}{\textit{98.26}} &\multicolumn{1}{c}{97.35} &\multicolumn{1}{c}{97.15} &\multicolumn{1}{|c}{\textit{0.89}} &\multicolumn{1}{c}{0.77} &\multicolumn{1}{c|}{1.26} \\
    \textbf{Fx} &\multicolumn{1}{|l|}{\cellcolor[HTML]{9dc8e3}\begin{tabular}[c]{@{}l@{}}ANN\\ (Quality driven)\end{tabular}} &\multicolumn{1}{c}{\cellcolor[HTML]{9dc8e3}\textit{98.25}} &\multicolumn{1}{c}{\cellcolor[HTML]{9dc8e3}97.30} &\multicolumn{1}{c}{\cellcolor[HTML]{9dc8e3}96.97} &\multicolumn{1}{|c}{\cellcolor[HTML]{9dc8e3}\textit{0.89}} &\multicolumn{1}{c}{\cellcolor[HTML]{9dc8e3}0.77} &\multicolumn{1}{c|}{\cellcolor[HTML]{9dc8e3}1.20} \\
      &\multicolumn{1}{|l|}{Polynomial} &\multicolumn{1}{c}{\textit{95.29}} &\multicolumn{1}{c}{96.06} &\multicolumn{1}{c}{89.19} &\multicolumn{1}{|c}{\textit{0.83}} &\multicolumn{1}{c}{0.83} &\multicolumn{1}{c|}{0.83} \\ \cline{2-8}
      &\multicolumn{1}{|l|}{\begin{tabular}[c]{@{}l@{}}ANN\\ (Bootstrap)\end{tabular}} &\multicolumn{1}{c}{\textit{98.04}} &\multicolumn{1}{c}{97.79} &\multicolumn{1}{c}{39.21} &\multicolumn{1}{|c}{\textit{0.73}} &\multicolumn{1}{c}{0.69} &\multicolumn{1}{c|}{1.06} \\
    \textbf{Fz} &\multicolumn{1}{|l|}{\cellcolor[HTML]{9dc8e3}\begin{tabular}[c]{@{}l@{}}ANN\\ (Quality driven)\end{tabular}} &\multicolumn{1}{c}{\cellcolor[HTML]{9dc8e3}\textit{97.98}} &\multicolumn{1}{c}{\cellcolor[HTML]{9dc8e3}97.78} &\multicolumn{1}{c}{\cellcolor[HTML]{9dc8e3}33.48} &\multicolumn{1}{|c}{\cellcolor[HTML]{9dc8e3}\textit{0.71}} &\multicolumn{1}{c}{\cellcolor[HTML]{9dc8e3}0.69} &\multicolumn{1}{c|}{\cellcolor[HTML]{9dc8e3}1.03} \\
      &\multicolumn{1}{|l|}{Polynomial} &\multicolumn{1}{c}{\textit{95.70}} &\multicolumn{1}{c}{97.54} &\multicolumn{1}{c}{75.85} &\multicolumn{1}{|c}{\textit{0.69}} &\multicolumn{1}{c}{0.69} &\multicolumn{1}{c|}{0.69} \\ \cline{2-8}
      &\multicolumn{1}{|l|}{\begin{tabular}[c]{@{}l@{}}ANN\\ (Bootstrap)\end{tabular}} &\multicolumn{1}{c}{\textit{99.95}} &\multicolumn{1}{c}{99.97} &\multicolumn{1}{c}{100.00} &\multicolumn{1}{|c}{\textit{0.06}} &\multicolumn{1}{c}{0.04} &\multicolumn{1}{c|}{0.08} \\
    \textbf{My} &\multicolumn{1}{|l|}{\cellcolor[HTML]{9dc8e3}\begin{tabular}[c]{@{}l@{}}ANN\\ (Quality driven)\end{tabular}} &\multicolumn{1}{c}{\cellcolor[HTML]{9dc8e3}\textit{99.79}} &\multicolumn{1}{c}{\cellcolor[HTML]{9dc8e3}99.84} &\multicolumn{1}{c}{\cellcolor[HTML]{9dc8e3}99.85} &\multicolumn{1}{|c}{\cellcolor[HTML]{9dc8e3}\textit{0.04}} &\multicolumn{1}{c}{\cellcolor[HTML]{9dc8e3}0.03} &\multicolumn{1}{c|}{\cellcolor[HTML]{9dc8e3}0.05} \\
      &\multicolumn{1}{|l|}{Polynomial} &\multicolumn{1}{c}{\textit{97.15}} &\multicolumn{1}{c}{97.31} &\multicolumn{1}{c}{98.92} &\multicolumn{1}{|c}{\textit{0.02}} &\multicolumn{1}{c}{0.02} &\multicolumn{1}{c|}{0.02} \\ \cline{2-8}
    \end{tabular}
    }
\end{table}

\begin{figure}[!t]
    \centering
    \includegraphics[width = \textwidth]{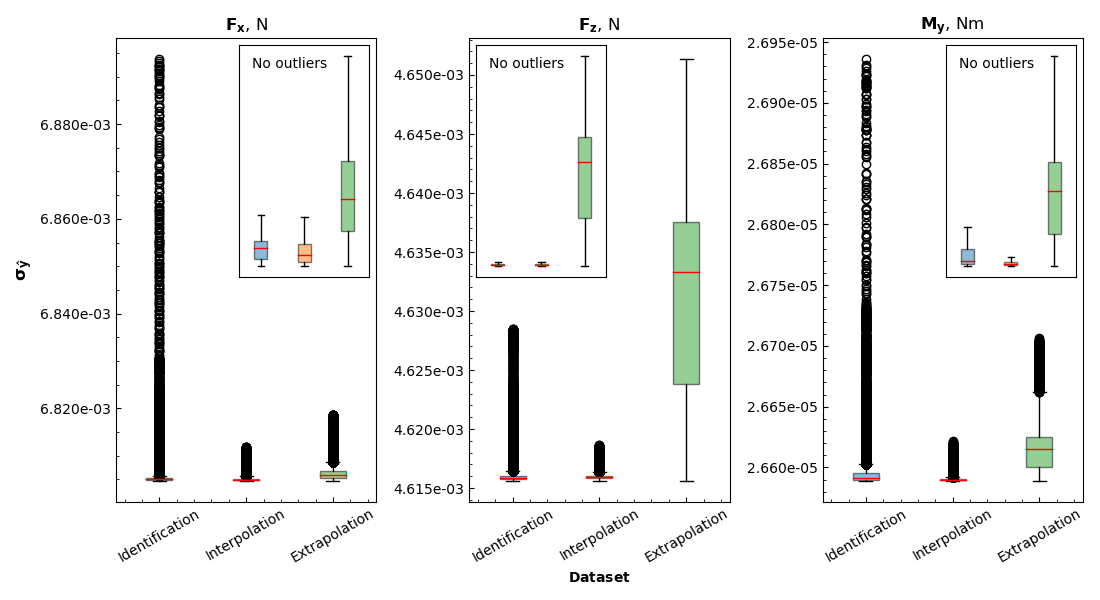}
    \caption{Comparison of the variance due to modelling uncertainty of the polynomial estimated PIs (second term in \cref{eq:OLS_sigma}) for the longitudinal models of the \textit{HDBeetle} for the identification, interpolation, and extrapolation data sets. The inset plots depict these comparisons without the outliers.}
    \label{fig:HDBeetle_PolyPIs_Boxplot}
\end{figure}

Both the bootstrap and quality driven ANN PI estimation techniques experience a decrease in MPIW when interpolating and an increase when extrapolating. This suggests that the ANN PIs adequately reflect uncertainty in the model due to extrapolation. As the estimated PIs are found to be valid in \cref{sec:numerical_validation}, they can be said to provide probabilistic bounds - which grow with uncertainty - within which the model predictions will likely lie despite their black-box nature. This is significant as it establishes an expected range of predictions for black-box models, subject to system uncertainties. Such information may be exploited to avoid surprises and inform (model-predictive or robust) controllers or to identify and evaluate extreme scenarios by considering the interval limits. Likewise, the PI growth can be monitored as a proxy for model extrapolation and confidence in ANN predictions. 


Contrarily, from \cref{tab:realPIPerf_InAndOutOfFlightEnvelope}, it would appear that the polynomial PI widths remain constant regardless of interpolation or extrapolation. This is inconsistent with the expected behaviour. Revisiting the polynomial PI equation, \cref{eq:OLS_sigma}, reveals that the PIs can only grow from differences between the observed model regressor variances and the new regressor values subject to the new input data. In fact, the polynomial PIs do indeed grow due to this model uncertainty term, albeit very slightly, as illustrated in \cref{fig:HDBeetle_PolyPIs_Boxplot}. It is clear that the extrapolation data set stimulates a significant widening of the polynomial PIs while the interpolation set does not. Note that the outliers represent 0.7\% of the data and may arise from artefacts in the data. For instance, in the wind tunnel, the local air speeds of the turbulent air near the edges of the nozzle are unknown. As the quadrotor is flown manually, excursions to these regions occur on occasion, especially for active flights. Likewise, un-modelled effects, such as reverse flow, which occur in the CyberZoo flights may also contribute to these outliers. While the growth of the polynomial PIs appears to be visually significant in \cref{fig:HDBeetle_PolyPIs_Boxplot}, the associated changes in magnitudes are minuscule in comparison to the PI width (for $F_{x}$: $\sigma_{\hat{y}} = 0.002\%$; $F_{z}$: $\sigma_{\hat{y}} = 0.048\%$; and $M_{y}$: $\sigma_{\hat{y}} = 0.070\%$ of MPIW). For the quadrotor models identified here (i.e. including the simulation models) it appears that the error residual variance, $\hat{\sigma}_{e}^{2}$ (first term in \cref{eq:OLS_sigma}), is the dominant term governing the width of the polynomial PI. Indeed, \cref{eq:OLS_sigma} is only valid for interpolation (i.e. within the scope of the training data) and not extrapolation as nothing can reliably be said about the new variation of the measurement data. This highlights some of the limitations of the Polynomial PI estimation technique, as there is - visually - minimal indication of extrapolation, unlike for the ANN counterparts. Nonetheless, this extrapolation is still apparent in the growth of the $\sigma_{\hat{y}}$ model uncertainty terms.

\begin{figure}[!t]
    \centering
    \includegraphics[width=\textwidth]{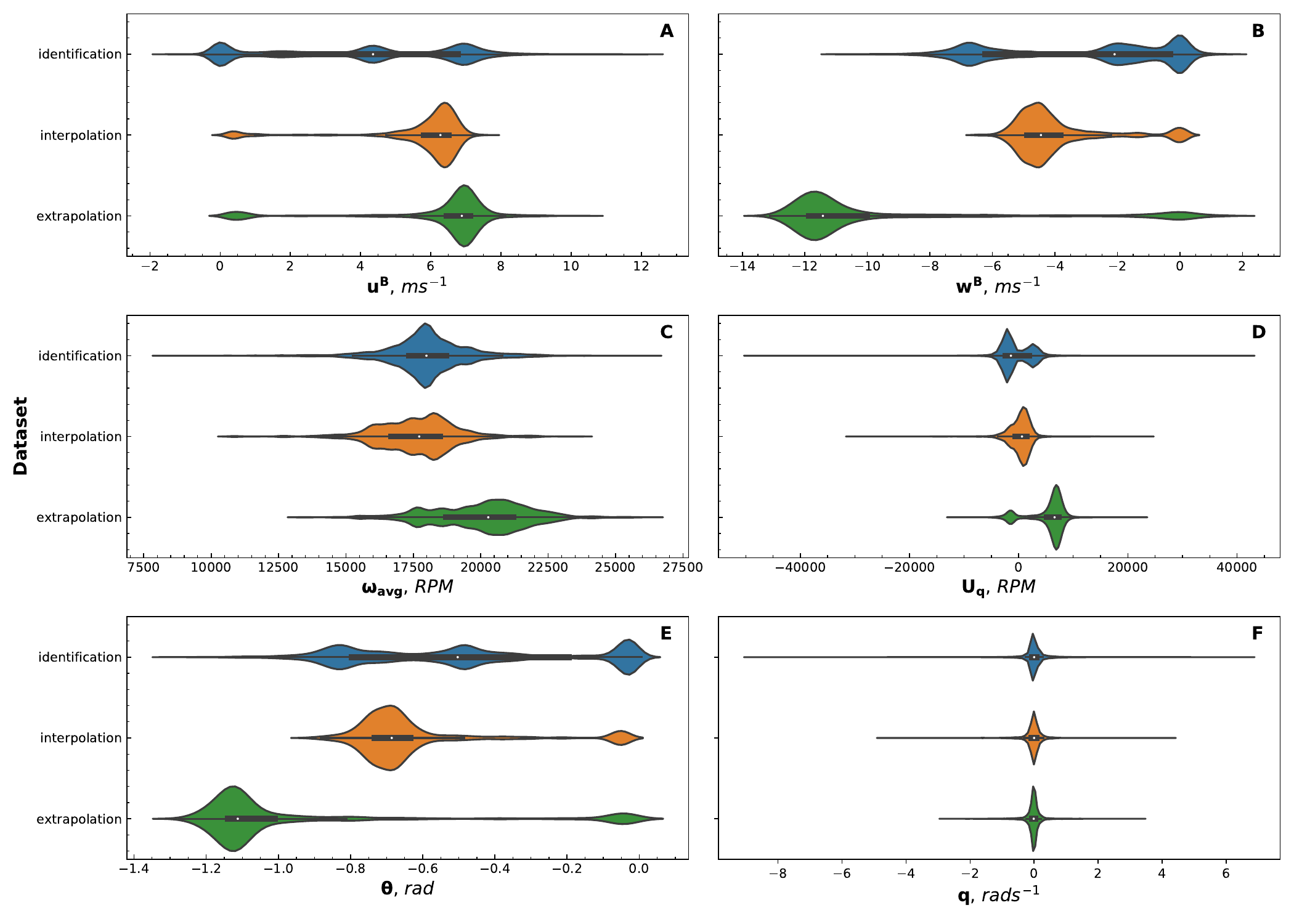}
    \caption{Differences in distribution, across the identification, interpolation, and extrapolation data sets, for key states of the longitudinal \textit{HDBeetle} models.}
    \label{fig:HDBeetle_DatasetDifferences_ViolinPlot}
\end{figure}

Furthermore, all PI estimation techniques show a drop in PICP when extrapolating, reflecting a degradation in model fit. This is most pronounced for models of $F_{z}$, and especially so for the ANN models. Overall, this deterioration in performance can be explained by differences between the identification, interpolation, and extrapolation datasets for key states of the longitudinal models, as summarized in \Cref{fig:HDBeetle_DatasetDifferences_ViolinPlot}. Indeed, the greatest differences between the extrapolation and identification sets occur for states related to $F_{z}$. Physically, the rotors need to work harder to counteract higher velocities (subplots \textbf{B} and \textbf{C} in \cref{fig:HDBeetle_DatasetDifferences_ViolinPlot}). This is only compounded by the increase in required trim pitch angle at higher velocities (subplot \textbf{E} in \cref{fig:HDBeetle_DatasetDifferences_ViolinPlot}). The comparatively similar state distributions of subplots \textbf{A}, \textbf{D}, and \textbf{F} account for the adequate PICP performance of the $F_{x}$ and $M_{y}$ models in \cref{tab:realPIPerf_InAndOutOfFlightEnvelope} and even afford valid PIs for the ANNs showcasing their well-known generalizability. Nonetheless, the fact that the models are extrapolating is evident through the MPIW.  



\section{Conclusion}\label{sec:conclusion}
Through mock data obtained from a high-fidelity quadrotor simulation, all considered prediction interval (PI) estimation techniques (polynomial, ANN bootstrap, and ANN quality driven) are shown to be numerically valid for aerodynamic quadrotor models subject to Gaussian noise statistics. Of the ANN estimation techniques, the bootstrap method appears to be superior in terms of PI quality and practicality. Indeed, the bootstrap method may be directly applied to an existing ensemble of ANN quadrotor models, provided that training datasets are available. Moreover, the polynomial PI results suggest that the width of the estimated PIs are reflective of the approximation power of the underlying model whereby excessively wide PIs may indicate a lack of approximation power. 

As the PI estimation techniques produce valid PIs, they are successfully applied to real quadrotor high-speed (up to $14$ $ms^{-1}$) flight data. The sensitivity of the estimated 95\% PIs to model interpolation and model extrapolation for each of the techniques is evaluated. Both ANN PI estimation techniques experience a swelling of PI widths for model extrapolation, reflecting a growing model uncertainty. These PIs remain unchanged, or shrink, when interpolating. Such behaviour is desirable in practice as insights into the model confidence and, crucially, the (95\%) prediction bounds of the otherwise black-box ANN models can be ascertained and accounted for. This behaviour is much less pronounced for the polynomial PIs, to the point that they are effectively unchanged for both model interpolation and extrapolation. The roots of this lie in the PI estimation itself, where the dominant term - associated with residual error variances - remains constant after training. Indeed, the model uncertainty term of the polynomial PI is reflective of model uncertainty but its magnitude is negligible in comparison to the measurement uncertainty term. This reveals a limitation to the analytically derived polynomial PIs when applied to model extrapolation. 

While the PIs discussed in this paper illustrate their utility, subsequent work should seek to expose further limitations of these PIs and investigate to what extent these PIs remain valid. For example, how do the estimated PIs react to changes in noise statistics - be this the distribution used, or changes to the Gaussian variances and means? Likewise, the relation between model complexity on the PI widths is also an interesting avenue for future work. Although challenging to do practically, it is desirable to also validate these PI estimation techniques using real flight data through repeated flights of a quadrotor in a controlled environment.



\appendix
\section{Appendix}\label{sec:app}

\subsection{Simulation Model Identification}\label{app:Sim_Models}

\subsubsection{Polynomial Candidate Regressors}\label{app:Sim_Polys_candRegs}

As these regressor pools can get quite large, the compact notation given in \cref{eq:PolyCands_Notation} is used to summarize the candidate pool. Here, $C_{0}$ denotes the bias vector, $P^{2}\left ( x, z \right )$ represents the polynomial basis, and $\left \{ 1, a \right \}$ the interacting factors. 

\begin{equation}\label{eq:PolyCands_Notation}
    \begin{array}{rl}
    \hat{Y} = & C_{0} + P^{2}\left ( x, z \right )\left \{ 1, a \right \} \\
    = & C_{0} + P^{2}\left ( x, z \right ) \cdot 1 +  P^{2}\left ( x, z \right ) \cdot a \\
     \text{where:} & \\
     \quad & P^{2}(x, z) = x^{2} + 2\cdot xz + z^{2}
    \end{array}
\end{equation}

The chosen candidate regressors for the simulation models are based on the gray-box models of Sun et al. \cite{Sun2018_GrayBox,Sun_ControlDoubleFailure}. In the following equations, terms outside the polynomials (i.e. $P^{d}(\cdot)$) are considered fixed regressors, chosen based on physical principles, and are thus always present in the model. 

The candidate force model structure for $F_{x}$, $F_{y}$, and $F_{z}$ are given by \cref{eq:simQuad_Structure_Fx}, \cref{eq:simQuad_Structure_Fy}, and \cref{eq:simQuad_Structure_Fz} respectively. 

\begin{flalign}\label{eq:simQuad_Structure_Fx}
    \begin{array}{rl}
    \hat{Fx} = & X_{0}\cdot \mu_{x} + P^3(\mu_{x},|\mu_{y}|,\mu_{z})\cdot \{1\}
    \end{array}
    &&
\end{flalign}

\begin{flalign}\label{eq:simQuad_Structure_Fy}
    \begin{array}{rl}
    \hat{Fy} = & Y_{0}\cdot \mu_{y} + P^3(|\mu_{x}|,\mu_{y},\mu_{z})\cdot \{1\}
    \end{array}
    &&
\end{flalign}

\begin{flalign}\label{eq:simQuad_Structure_Fz}
    \begin{array}{rl}
    \hat{Fz} = & Z_{0}\cdot ({\mu_{x}^{2} + \mu_{y}^{2}}) - Z_{1} {\cdot (\mu_{z} - \mu_{v_{in}}}) + P^3(|\mu_{x}|,|\mu_{y}|,\mu_{z})\cdot \{1\} + P^3(|U_{p}|,|U_{q}|,U_{r})\cdot \{1,\omega_{avg}\}\\
    & + P^3(|p|,|q|,r)\cdot \{1,\omega_{avg}\} + P^2(\omega_{avg})\cdot \{1\}\\
    \end{array}
    &&
\end{flalign}

The candidate moment model structure for $M_{x}$, $M_{y}$, and $M_{z}$ are given by \cref{eq:simQuad_Structure_Mx}, \cref{eq:simQuad_Structure_My}, and \cref{eq:simQuad_Structure_Mz} respectively. 

\begin{flalign}\label{eq:simQuad_Structure_Mx}
    \begin{array}{rl}
    \hat{Mx} = & L_{0}\cdot U_{p} + P^3(\mu_{y},\mu_{z})\cdot \{1,p,U_{p}\} + P^3(p,U_{p})\cdot \{1,\omega_{avg}\}\\
    \end{array}
    &&
\end{flalign}

\begin{flalign}\label{eq:simQuad_Structure_My}
    \begin{array}{rl}
    \hat{My} = & M_{0}\cdot U_{q} + P^3(\mu_{x},\mu_{z})\cdot \{1,q,U_{q}\} + P^3(q,U_{q})\cdot \{1,\omega_{avg}\}\\
    \end{array}
    &&
\end{flalign}

\begin{flalign}\label{eq:simQuad_Structure_Mz}
    \begin{array}{rl}
    \hat{Mz} = & N_{0}\cdot U_{r} + P^3(\mu_{x},\mu_{y},\mu_{z})\cdot \{1,|p|,|q|,r,|U_{p}|,|U_{q}|,U_{r},\omega_{avg}\} + P^3(|p|,|q|,r)\cdot \{1,\omega_{avg}\}\\
    & + P^3(|U_{p}|,|U_{q}|,U_{r})\cdot \{1,\omega_{avg}\}\\
    \end{array}
    &&
\end{flalign}

\subsubsection{ANN Input Vectors}\label{app:Sim_ANN}

\begin{table}[H]
    \caption{Input vectors employed for the simulation model identification for both the ANN bootstrap and ANN quality driven techniques.}
    \label{tab:simModel_ANN_Inputs}
    \centering
    \begin{tabular}{rl}\textbf{Model} &\textbf{Input variables} \\ \hline
    \multicolumn{1}{r|}{Fx} &$\mu_{x}$, $|\mu_{y}|$, $\mu_{z}$ \\
    \multicolumn{1}{r|}{Fy} &$|\mu_{x}|$, $\mu_{y}$, $\mu_{z}$ \\
    \multicolumn{1}{r|}{Fz} &$w$, $|\mu_{x}|$, $|\mu_{y}|$, $\mu_{z}$, $\omega_{avg}$, $\mu_{v_{in}}$, $|U_{p}|$, $|U_{q}|$, $U_{r}$, $|p|$, $|q|$, $r$ \\
    \multicolumn{1}{r|}{Mx} &$\mu_{y}$, $\mu_{z}$, $U_{p}$, $p$, $\omega_{avg}$ \\
    \multicolumn{1}{r|}{My} &$\mu_{x}$, $\mu_{z}$, $U_{q}$, $q$, $\omega_{avg}$ \\
    \multicolumn{1}{r|}{Mz} &$\mu_{x}$, $\mu_{y}$, $\mu_{z}$, $p$, $q$, $r$, $U_{p}$, $U_{q}$, $U_{r}$, $\omega_{avg}$ \\
    \hline\end{tabular}
\end{table}

\subsubsection{Identified Polynomial Model Structures}\label{app:Sim_Polys_modelStruct}

\begin{table}[!h]
    \centering
    \caption{Identified polynomial force model structures, coefficients, and goodness-of-fit metrics for the simulation model. The rows indicate the order of addition to the model, as selected by the stepwise regression algorithm. Rows in grey denote fixed regressors.}
    \label{tab:simModel_ForceStructures}
    \begin{tabular}{ccc|ccc|ccc}
    \multicolumn{3}{c}{\textbf{Fx}} &\multicolumn{3}{c}{\textbf{Fy}} &\multicolumn{3}{c}{\textbf{Fz}} \\ \hline 
    Regressor &Coefficient &R2 &Regressor &Coefficient &R2 &Regressor &Coefficient &R2 \\ \hline 
    \cellcolor[HTML]{DCDCDC}$bias$ &\cellcolor[HTML]{DCDCDC}-1.431e-04 &\cellcolor[HTML]{DCDCDC}- &\cellcolor[HTML]{DCDCDC}$bias$ &\cellcolor[HTML]{DCDCDC}-2.187e-05 &\cellcolor[HTML]{DCDCDC}- &\cellcolor[HTML]{DCDCDC}$bias$ &\cellcolor[HTML]{DCDCDC}-1.654e+00 &\cellcolor[HTML]{DCDCDC}- \\
    \cellcolor[HTML]{DCDCDC}$\mu_{x}$ &\cellcolor[HTML]{DCDCDC}-2.498e-01 &\cellcolor[HTML]{DCDCDC}0.98 &\cellcolor[HTML]{DCDCDC}$\mu_{y}$ &\cellcolor[HTML]{DCDCDC}-2.675e-01 &\cellcolor[HTML]{DCDCDC}0.97 &\cellcolor[HTML]{DCDCDC}${\mu_{x}^{2} + \mu_{y}^{2}}$ &\cellcolor[HTML]{DCDCDC}3.209e-02 &\cellcolor[HTML]{DCDCDC}- \\
     ${\mu_{x}^{1.0} \cdot  \mu_{z}^{1.0}}$ & 1.780e-02 & 0.98 & ${\mu_{y}^{1.0} \cdot  \mu_{z}^{1.0}}$ & 2.779e-02 & 0.98 &\cellcolor[HTML]{DCDCDC}${-\mu_{z} + \mu_{v_{in}}}$ &\cellcolor[HTML]{DCDCDC}-1.070e-01 &\cellcolor[HTML]{DCDCDC}0.01 \\
     $-$ & - & - & ${|\mu_{x}|^{1.0} \cdot  \mu_{y}^{1.0}}$ & -1.844e-02 & 0.98 & ${\omega_{avg}^{2.0}}$ & -1.334e-02 & 0.85 \\
     $-$ & - & - & $-$ & - & - & ${\omega_{avg}}\cdot {U_{r}^{2.0}}$ & -1.603e-08 & 0.87 \\
     $-$ & - & - & $-$ & - & - & ${\omega_{avg}^{1.0}}$ & 2.746e-01 & 0.88 \\
     $-$ & - & - & $-$ & - & - & ${\mu_{z}^{1.0}}$ & -5.148e-01 & 0.89 \\
     $-$ & - & - & $-$ & - & - & ${\omega_{avg}}\cdot {|U_{q}|^{3.0}}$ & -1.218e-11 & 0.89 \\
    \hline
    \end{tabular}
\end{table}

\begin{table}[!h]
    \centering
    \caption{Identified polynomial moment model structures, coefficients, and goodness-of-fit metrics for the simulation model. The rows indicate the order of addition to the model, as selected by the stepwise regression algorithm. Rows in grey denote fixed regressors.}
    \label{tab:simModel_MomentStructures}
    \begin{tabular}{ccc|ccc|ccc}
    \multicolumn{3}{c}{\textbf{Mx}} &\multicolumn{3}{c}{\textbf{My}} &\multicolumn{3}{c}{\textbf{Mz}} \\ \hline 
    Regressor &Coefficient &R2 &Regressor &Coefficient &R2 &Regressor &Coefficient &R2 \\ \hline 
    \cellcolor[HTML]{DCDCDC}$bias$ &\cellcolor[HTML]{DCDCDC}-3.838e-05 &\cellcolor[HTML]{DCDCDC}- &\cellcolor[HTML]{DCDCDC}$bias$ &\cellcolor[HTML]{DCDCDC}5.530e-05 &\cellcolor[HTML]{DCDCDC}- &\cellcolor[HTML]{DCDCDC}$bias$ &\cellcolor[HTML]{DCDCDC}1.705e-05 &\cellcolor[HTML]{DCDCDC}- \\
    \cellcolor[HTML]{DCDCDC}$U_{p}$ &\cellcolor[HTML]{DCDCDC}2.938e-04 &\cellcolor[HTML]{DCDCDC}0.89 &\cellcolor[HTML]{DCDCDC}$U_{q}$ &\cellcolor[HTML]{DCDCDC}2.214e-04 &\cellcolor[HTML]{DCDCDC}0.85 &\cellcolor[HTML]{DCDCDC}$U_{r}$ &\cellcolor[HTML]{DCDCDC}-3.321e-06 &\cellcolor[HTML]{DCDCDC}0.94 \\
     ${\mu_{y}^{1.0}}$ & -2.649e-02 & 0.99 & ${\mu_{x}^{1.0}}$ & 2.950e-02 & 0.98 & ${r^{1.0}}$ & -2.295e-03 & 0.97 \\
     ${U_{p}}\cdot {\mu_{y}^{2.0}}$ & -5.857e-06 & 0.99 & ${\mu_{x}^{1.0} \cdot  \mu_{z}^{1.0}}$ & 1.036e-02 & 0.99 & ${|U_{q}|}\cdot {\mu_{x}^{1.0} \cdot  \mu_{y}^{1.0} \cdot  \mu_{z}^{1.0}}$ & 1.981e-06 & 0.98 \\
     $-$ & - & - & ${U_{q}}\cdot {\mu_{z}^{2.0}}$ & -1.298e-05 & 0.99 & ${\omega_{avg}}\cdot {U_{r}^{1.0}}$ & 1.203e-06 & 0.98 \\
    \hline
    \end{tabular}
\end{table}

\subsection{Gaussian Noise Statistics}\label{app:gaussianNoiseStats}

\begin{table}[H]
    \centering
    \caption{Gaussian noise statistics used to contaminate the simulation model inputs and measurements with uncertainty. The noise standard deviations are selected to be of the same magnitude as, but slightly worse than, commonly used sensors such as the MPU6000 \cite{gonzalez2019MPU600performance}.}
    \label{tab:noiseStatistics_states}
    \resizebox{0.95\textwidth}{!}{%
    \begin{tabular}{l|r|rrr|rrr|rrr|rrr|}
    \cline{2-14}
                                                      & \textbf{\begin{tabular}[c]{@{}l@{}}Rotor \\ speeds {[}eRPM{]}\end{tabular}} & \multicolumn{3}{l|}{\textbf{Velocity {[}ms$^{-1}${]}}} & \multicolumn{3}{l|}{\textbf{Acceleration {[}ms$^{-2}${]}}} & \multicolumn{3}{l|}{\textbf{Attitude {[}rad{]}}} & \multicolumn{3}{l|}{\textbf{Body rates {[}rads$^{-1}${]}}} \\
                                                      & $\Omega$                  & \textit{u}      & \textit{v}     & \textit{w}     & $a_{x}$      & $a_{y}$      & $a_{z}$     & $\phi$  & $\theta$  & $\psi$  & \textit{p}      & \textit{q}      & \textit{r}      \\ \cline{1-1} \hline
    \multicolumn{1}{|l|}{\textbf{Mean}, $\mu$}               & 0                                                                           & 0               & 0              & 0              & 0                & 0                & 0               & 0              & 0               & 0             & 0               & 0               & 0               \\
    \multicolumn{1}{|l|}{\textbf{Standard deviation}, $\sigma$} & 5                                                                           & 0.03            & 0.03           & 0.03           & 0.03             & 0.03             & 0.03            & 0.01           & 0.01            & 0.01          & 0.01            & 0.01            & 0.01            \\ \hline
    \end{tabular}
    }
\end{table}

\begin{figure}[H]
    \centering
    \includegraphics[width = 0.95\textwidth]{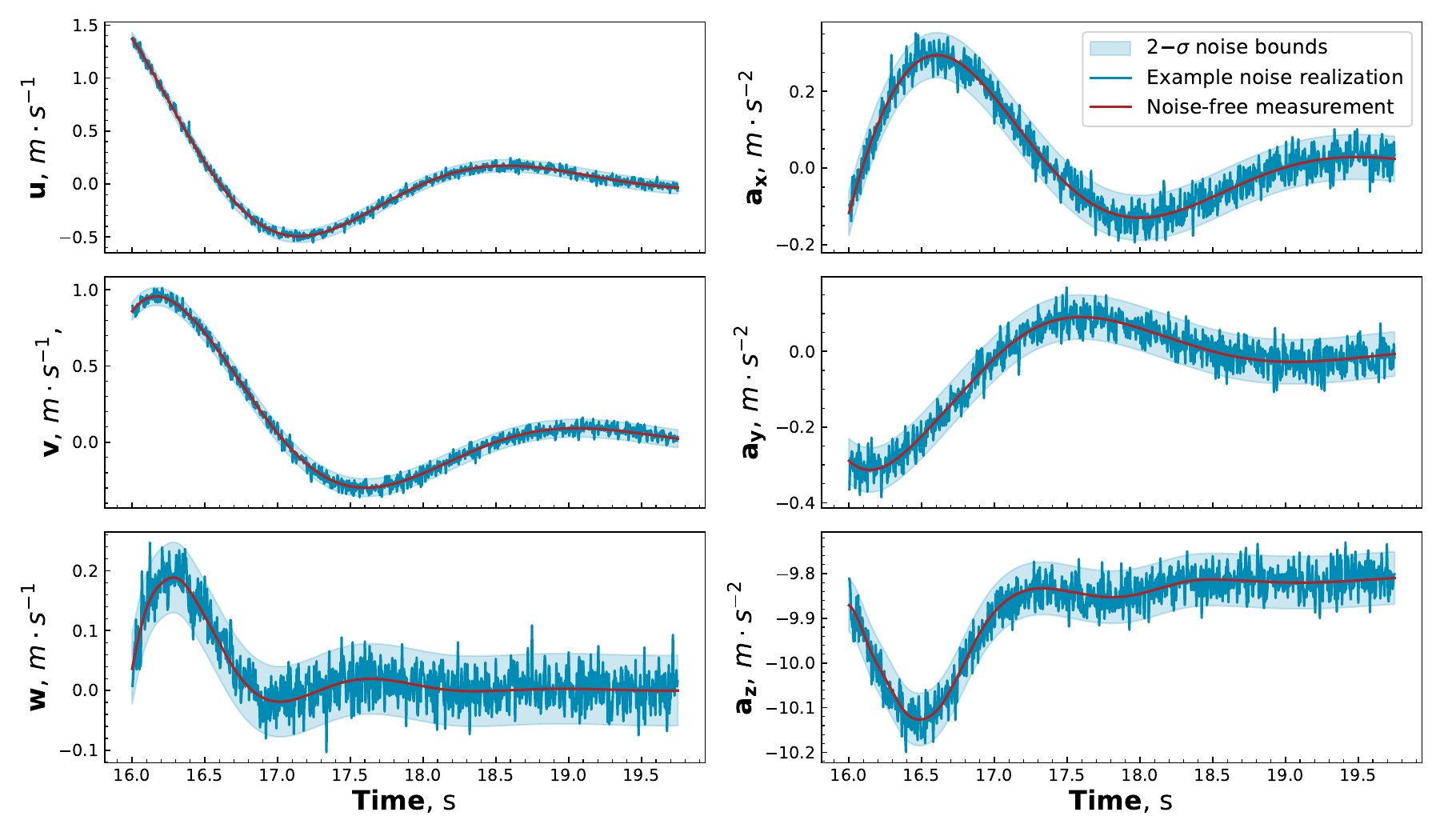}
    \caption{Illustration of the addition of Gaussian noise to the noise free measurements (red) of the velocities and accelerations to produce associated noisy realizations (blue). The shaded region represents the 2$-\sigma$ bounds of the noise statistics used.}
    \label{fig:ExampleNoise_Velocities_Accelerations}
\end{figure}

\begin{figure}[H]
    \centering
    \includegraphics[width = 0.95\textwidth]{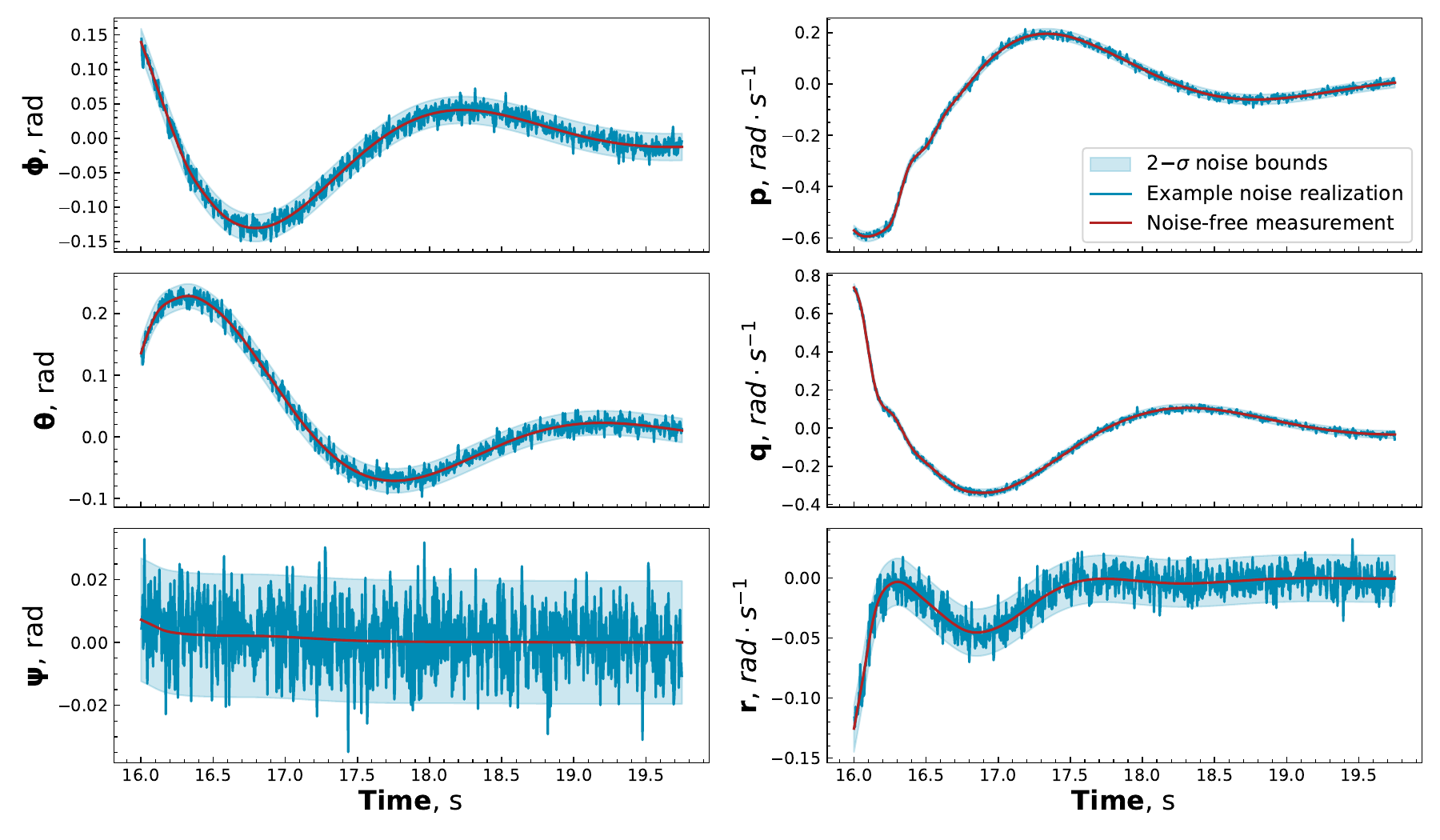}
    \caption{Illustration of the addition of Gaussian noise to the noise free measurements (red) of the attitude and rotational rates to produce associated noisy realizations (blue). The shaded region represents the 2$-\sigma$ bounds of the noise statistics used.}
    \label{fig:ExampleNoise_attitude_rates}
\end{figure}

\begin{figure}[H]
    \centering
    \includegraphics[width = 0.95\textwidth]{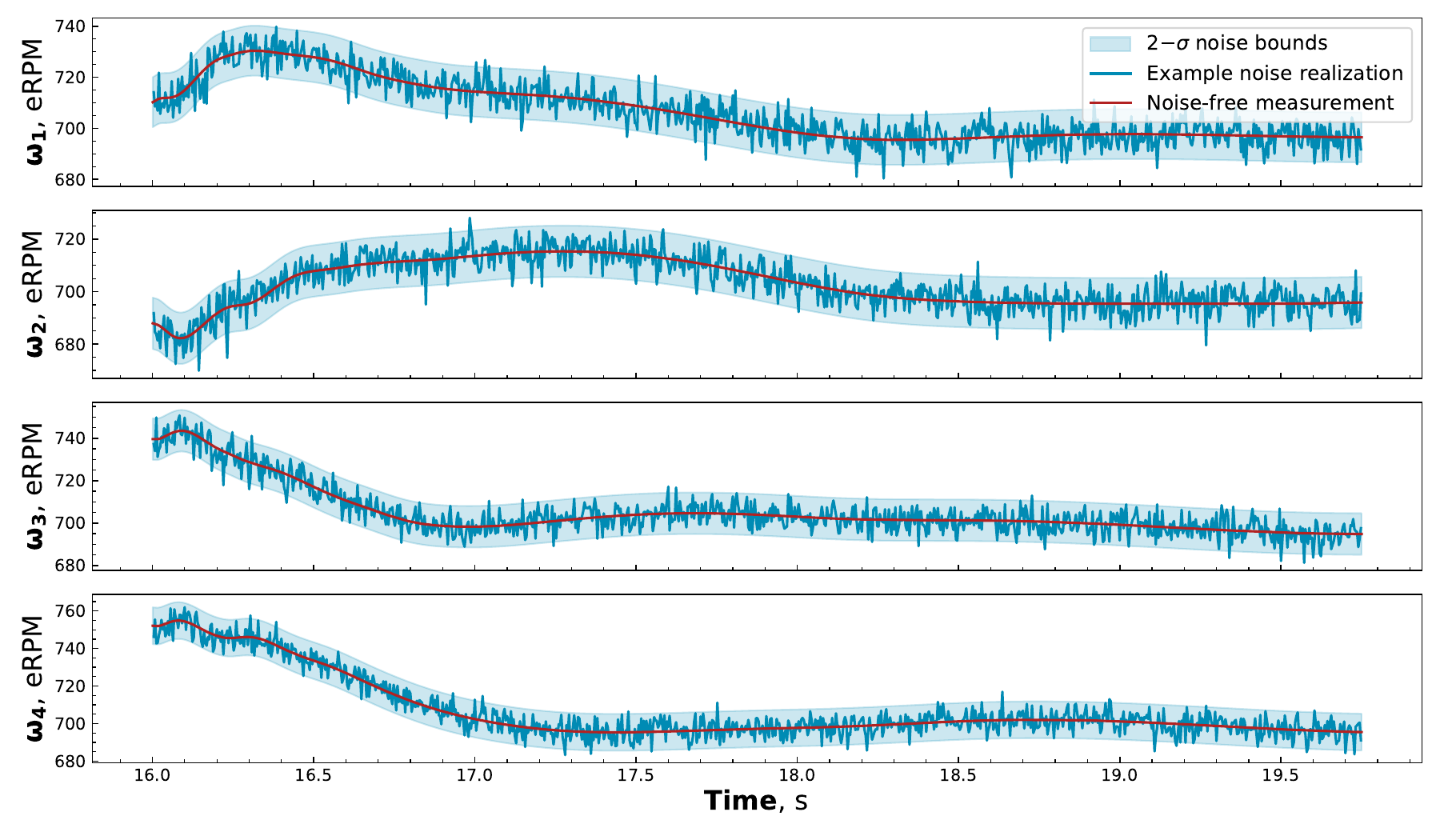}
    \caption{Illustration of the addition of Gaussian noise to the noise free measurements (red) of the rotor speeds to produce associated noisy realizations (blue). The shaded region represents the 2$-\sigma$ bounds of the noise statistics used.}
    \label{fig:ExampleNoise_rotorSpeeds}
\end{figure}

\subsection{HDBeetle Quadrotor Model Identification}\label{app:Real_Model}

\subsubsection{Polynomial Candidate Regressors}\label{app:realPolys_candidateRegressors}

The candidate regressors for the longitudinal \textit{HDBeetle} models are chosen based on the grey-box models of Sun et al. \cite{Sun2018_GrayBox,Sun_ControlDoubleFailure} and from experimentation. The candidate regressor structures for $F_{x}$, $F_{z}$, and $M_{y}$ are summarized in \cref{eq:realQuad_Structure_Fx}, \cref{eq:realQuad_Structure_Fz}, and \cref{eq:realQuad_Structure_My} respectively. 

\begin{flalign}\label{eq:realQuad_Structure_Fx}
    \begin{array}{rl}
    \hat{F}_{x} = & X_{0}\cdot u + X_{1}\cdot \sin{{\theta}} + P^3(u,|v|,w)\cdot \{1,\omega_{avg},q,U_{q},\sin{{\theta}},\cos{{\theta}}\}\\
    & + P^3(q,|r|)\cdot \{1,\omega_{avg}\} + P^3(\mu_{x},|\mu_{y}|,\mu_{z})\cdot \{1,\omega_{avg},q,U_{q},\sin{{\theta}},\cos{{\theta}}\}\\
    & + P^3(U_{q},|U_{p}|,|U_{r}|)\cdot \{1,\sin{{\theta}},\cos{{\theta}},\omega_{avg}\} + P^2(\omega_{avg})\cdot \{\sin{{\theta}}\}\\
    \end{array}
    &&
\end{flalign}

\begin{flalign}\label{eq:realQuad_Structure_Fz}
    \begin{array}{rl}
    \hat{F}_{z} = & Z_{0}\cdot w + Z_{1}\cdot \cos{{\theta}}\cdot \cos{{\phi}}\\
    & + P^3(|u|,|v|,w)\cdot \{1,\omega_{avg},|p|,|q|,|r|,|U_{p}|,|U_{q}|,|U_{r}|,\sin{{\theta}},\cos{{\theta}},\sin{{\phi}},\cos{{\phi}}\}\\
    & + P^3(|p|,|q|,|r|)\cdot \{1,\omega_{avg}\}\\
    & + P^3(|\mu_{x}|,|\mu_{y}|,\mu_{z})\cdot \{1,\omega_{avg},|p|,|q|,|r|,|U_{p}|,|U_{q}|,|U_{r}|,\sin{{\theta}},\cos{{\theta}},\sin{{\phi}},\cos{{\phi}}\}\\
    & + P^3(|U_{q}|,|U_{p}|,|U_{r}|)\cdot \{1,\sin{{\psi}},\cos{{\psi}},\omega_{avg}\} + P^4(\omega_{avg})\cdot \{1,\sin{{\theta}},\cos{{\theta}},\sin{{\phi}},\cos{{\phi}}\}\\
    \end{array}
    &&
\end{flalign}

\begin{flalign}\label{eq:realQuad_Structure_My}
    \begin{array}{rl}
    \hat{My} = & M_{0}\cdot q + M_{1}\cdot U_{q} + P^4(u,|v|,w)\cdot \{1,\omega_{avg},q,U_{q},\sin{{\theta}},\cos{{\theta}}\}\\
    & + P^4(|p|,q,|r|)\cdot \{1,\omega_{avg},\sin{{\theta}},\cos{{\theta}}\} + P^4(\mu_{x},|\mu_{y}|,\mu_{z})\cdot \{1,\omega_{avg},q,U_{q},\sin{{\theta}},\cos{{\theta}}\}\\
    & + P^4(U_{q},|U_{p}|,|U_{r}|)\cdot \{1,\sin{{\theta}},\cos{{\theta}},\omega_{avg}\} + P^4(\omega_{avg})\cdot \{\sin{{\theta}},\cos{{\theta}}\}\\
    \end{array}
    &&
\end{flalign}

\subsubsection{ANN Input Vectors}\label{app:sub_Real_ANN}

\begin{table}[!h]
    \centering
    \caption{Input vectors employed for the \textit{HDBeetle} longitudinal model identification for both the ANN bootstrap and ANN quality driven techniques.}
    \label{tab:realModel_ANN_Inputs}
    \begin{tabular}{rl}\textbf{Model} &\textbf{Input variables} \\ \hline
    \multicolumn{1}{r|}{Fx} &$u$, $|v|$, $w$, $q$, $|r|$, $\omega_{avg}$, $U_{q}$, $\sin{{\theta}}$, $\cos{{\theta}}$, $\mu_{x}$, $|\mu_{y}|$, $\mu_{z}$, $|U_{r}|$, $|U_{p}|$ \\
    \multicolumn{1}{r|}{Fz} &$|u|$, $|v|$, $w$, $\omega_{avg}$, $|p|$, $|q|$, $|r|$, $|U_{p}|$, $|U_{q}|$, $|U_{r}|$, $\sin{{\theta}}$, $\cos{{\theta}}$, $\sin{{\phi}}$, $\cos{{\phi}}$, $|\mu_{x}|$, $|\mu_{y}|$, $\mu_{z}$ \\
    \multicolumn{1}{r|}{My} &$u$, $|v|$, $w$, $\omega_{avg}$, $|p|$, $q$, $|r|$, $U_{q}$, $\sin{{\theta}}$, $\cos{{\theta}}$, $\mu_{x}$, $|\mu_{y}|$, $\mu_{z}$, $|U_{p}|$, $|U_{r}|$ \\
    \hline\end{tabular}
\end{table}

\subsubsection{Identified Polynomial Model Structures}\label{app:real_modelstructure}

\begin{table}[H]
    \centering
    \caption{Identified polynomial longitudinal model structures, coefficients, and goodness-of-fit metrics for the \textit{HDBeetle}. The rows indicate the order of addition to the model, as selected by the stepwise regression algorithm. Rows in grey denote fixed regressors.}
    \label{tab:HDBeetleModel_MomentStructures}
    \begin{tabular}{ccc|ccc|ccc}
    \multicolumn{3}{c}{\textbf{Fx}} &\multicolumn{3}{c}{\textbf{Fz}} &\multicolumn{3}{c}{\textbf{My}} \\ \hline 
    Regressor &Coefficient &R2 &Regressor &Coefficient &R2 &Regressor &Coefficient &R2 \\ \hline 
    \cellcolor[HTML]{DCDCDC}$bias$ &\cellcolor[HTML]{DCDCDC}-7.542e-02 &\cellcolor[HTML]{DCDCDC}- &\cellcolor[HTML]{DCDCDC}$bias$ &\cellcolor[HTML]{DCDCDC}2.413e-01 &\cellcolor[HTML]{DCDCDC}- &\cellcolor[HTML]{DCDCDC}$bias$ &\cellcolor[HTML]{DCDCDC}6.892e-03 &\cellcolor[HTML]{DCDCDC}- \\
    \cellcolor[HTML]{DCDCDC}$u$ &\cellcolor[HTML]{DCDCDC}2.884e-02 &\cellcolor[HTML]{DCDCDC}- &\cellcolor[HTML]{DCDCDC}$w$ &\cellcolor[HTML]{DCDCDC}-7.440e-02 &\cellcolor[HTML]{DCDCDC}- &\cellcolor[HTML]{DCDCDC}$q$ &\cellcolor[HTML]{DCDCDC}7.872e-04 &\cellcolor[HTML]{DCDCDC}- \\
    \cellcolor[HTML]{DCDCDC}$\sin{{\theta}}$ &\cellcolor[HTML]{DCDCDC}-1.398e-01 &\cellcolor[HTML]{DCDCDC}0.93 &\cellcolor[HTML]{DCDCDC}$\cos{{\theta}}\cdot \cos{{\phi}}$ &\cellcolor[HTML]{DCDCDC}-6.245e-01 &\cellcolor[HTML]{DCDCDC}0.45 &\cellcolor[HTML]{DCDCDC}$U_{q}$ &\cellcolor[HTML]{DCDCDC}8.010e-04 &\cellcolor[HTML]{DCDCDC}0.73 \\
     ${\omega_{avg}}\cdot {u^{1.0}}$ & -4.537e-03 & 0.96 & ${\omega_{avg}^{2.0}}$ & -2.946e-04 & 0.95 & ${\omega_{avg}}\cdot {\mu_{z}^{2.0}}$ & -7.295e-06 & 0.92 \\
     ${\cos{{\theta}}}\cdot {u^{3.0}}$ & 1.636e-03 & 0.97 & ${\sin{{\theta}}}\cdot {|u|^{1.0} \cdot  w^{2.0}}$ & -3.229e-03 & 0.96 & ${\omega_{avg}}\cdot {\mu_{x}^{1.0} \cdot  \mu_{z}^{1.0}}$ & 1.338e-04 & 0.92 \\
     $-$ & - & - & $-$ & - & - & ${\mu_{x}^{1.0}}$ & 1.217e-02 & 0.94 \\
    \hline
    \end{tabular}
\end{table}

\bibliography{sample.bib}


\end{document}